\documentclass[preprint2]{aastex62}
\usepackage{natbib}
\usepackage{hyperref}
\usepackage{breakurl}
\bibpunct[; ]{(}{)}{,}{a}{}{;}
\submitjournal{ApJ}
\received{}
\revised{}
\accepted{}

\usepackage{color}

\shorttitle{Spectral content in SDO/AIA 1600 and 1700 \AA\ filters}
\shortauthors{Sim\~oes et al.}

\begin{document}

\title{The spectral content of SDO/AIA 1600 and 1700~\AA~ filters from flare and plage observations}

\correspondingauthor{Paulo J. A. Sim\~oes}
\email{paulo.simoes@glasgow.ac.uk}

\author[0000-0002-4819-1884]{Paulo J. A. Sim\~oes}
\affiliation{SUPA School of Physics and Astronomy, University of Glasgow, Glasgow, G12 8QQ, UK}
\affiliation{Centro de R\'adio Astronomia e Astrof\'isica Mackenzie, CRAAM, Universidade Presbiteriana Mackenzie, S\~ao Paulo, SP 01302-907, Brazil}

\author[0000-0002-6287-3494]{Hamish A. S. Reid}
\affiliation{SUPA School of Physics and Astronomy, University of Glasgow, Glasgow, G12 8QQ, UK}

\author[0000-0001-5031-1892]{Ryan O. Milligan}
\affiliation{SUPA School of Physics and Astronomy, University of Glasgow, Glasgow, G12 8QQ, UK}
\affiliation{Solar Physics Laboratory (Code 671), Heliophysics Science Division, NASA Goddard Space Flight Center, Greenbelt, MD 20771, USA}
\affiliation{Department of Physics Catholic University of America, 620 Michigan Avenue, Northeast, Washington, DC 20064, USA}

\author[0000-0001-9315-7899]{Lyndsay Fletcher}
\affiliation{SUPA School of Physics and Astronomy, University of Glasgow, Glasgow, G12 8QQ, UK}

\begin{abstract}
The strong enhancement of the ultraviolet emission during solar flares is usually taken as an indication of plasma heating in the lower solar atmosphere caused by the deposition of the energy released during these events. Images taken with broadband ultraviolet filters by the {\em Transition Region and Coronal Explorer} (TRACE) and {\em Atmospheric Imaging Assembly} (AIA 1600 and 1700~\AA) have revealed the morphology and evolution of flare ribbons in great detail. However, the spectral content of these images is still largely unknown. Without the knowledge of the spectral contribution to these UV filters, the use of these rich imaging datasets is severely limited. Aiming to solve this issue, we estimate the spectral contributions of the AIA UV flare and plage images using high-resolution spectra in the range 1300 to 1900~\AA\ from the Skylab NRL SO82B spectrograph. We find that the flare excess emission in AIA 1600~\AA\ is { dominated by} the \ion{C}{4} 1550~\AA\ doublet (26\%), \ion{Si}{1} continua (20\%), with smaller contributions from many other chromospheric lines such as \ion{C}{1} 1561 and 1656~\AA\ multiplets, \ion{He}{2} 1640~\AA, \ion{Si}{2} 1526 and 1533~\AA. For the AIA 1700~\AA\ band, \ion{C}{1} 1656~\AA\ multiplet is the main contributor (38\%), followed by \ion{He}{2} 1640 (17\%), and accompanied by a multitude of other, { weaker} chromospheric lines, with minimal contribution from the continuum. Our results can be generalized to state that the AIA UV flare excess emission is of chromospheric origin, while plage emission is dominated by photospheric continuum emission in both channels. 

\end{abstract}


\section{Introduction} \label{sec:intro}

Uultraviolet (UV) emission {from solar flares  in the 1000--2000~\AA\ range} is often regarded as evidence of plasma heating in the lower solar atmosphere, highlighting the location of the energy deposition, { which is often coincident with} hard X-ray emission \citep[e.g.][]{WarrenWarshall:2001,AlexanderCoyner:2006,FletcherHudson:2001,SimoesGrahamFletcher:2015a} and mapping the connectivity of the flaring magnetic loops \citep[e.g.][]{Fletcher:2009,JoshiVeronigCho:2009,ReidVilmerAulanier:2012}.
UV imaging of solar flares by the {\em Transition Region and Coronal Explorer} \citep[TRACE;][]{HandyActonKankelborg:1999} and the {\em Atmospheric Imaging Assembly} \citep[AIA;][]{LemenTitleAkin:2012} on board the {\em Solar Dynamics Observatory} \cite[SDO;][]{PesnellThompsonChamberlin:2012} has been used extensively to investigate the spatial morphology and temporal evolution of flare ribbons \citep[e.g.][]{WarrenWarshall:2001,FletcherPollockPotts:2004,SimoesFletcherHudson:2013,KazachenkoLynchWelsch:2017}. 

{ Due to a lack of knowledge of the spectral content of the AIA UV images, detailed qualitative analyses of flaring plasma using AIA UV images have been rare.} A few exceptions are the investigation of plasma cooling \citep{QiuLiuHill:2010,ChengKerrQiu:2012,QiuSturrockLongcope:2013} and estimates of the radiative output { from the flaring chromosphere} \citep{MilliganKerrDennis:2014}. In contrast, the spectral content in the AIA extreme ultraviolet (EUV) images has been extensively explored. The radiation of the vast majority of spectral lines in the EUV range can be assumed to be optically thin. This assumption greatly facilitates the analysis of EUV data by the use of synthetic spectra generated by CHIANTI \citep{DereLandiMason:1997,LandiYoungDere:2013}, which has led to significant advancements in understanding the evolution of hot plasmas in the Sun 
\citep[e.g.][]{YoungDel-ZannaMason:2007,MilliganChamberlinHudson:2012,Del-ZannaWoods:2013}, and identifying the response of the EUV filters on AIA \citep{ODwyerDel-ZannaMason:2010,MilliganMcElroy:2013}. This knowledge of the spectral response of the AIA EUV channels has led to the development of techniques to estimate the differential emission measure of the plasma \citep[e.g.][]{HannahKontar:2012,Del-Zanna:2013,CheungBoernerSchrijver:2015}, and several other studies of the physical characteristics of the plasma in active regions \citep[e.g.][]{SchmelzJenkinsWorley:2011,Del-Zanna:2013}, microflares \citep[e.g.][]{InglisChriste:2014,WrightHannahGrefenstette:2017} and flares \citep[e.g.][]{HannahKontar:2013,BattagliaFletcherSimoes:2014,SimoesGrahamFletcher:2015a}. In the UV range, however, the presence of recombination edges and spectral lines originating from neutral and weakly-ionized metals, which are unlikely to be optically thin, prevents the direct use of synthetic spectra for similar analysis.

One of the purposes of the UV filters on TRACE ({ 1550, 1600 and 1700~\AA}) was to estimate the \ion{C}{4} emission \citep{HandyBrunerTarbell:1998}, given the expected radiative importance of this line \citep[e.g.][]{BrunerMcWhirter:1988,HawleyFisher:1994}, and the potential of using its flux as a transition region `pressure gauge' \citep{HawleyFisher:1992}. TRACE \ion{C}{4} images have been explored for studies of quiescent structures \citep{de-WijnDe-Pontieu:2006,De-PontieuTarbellErdelyi:2003}, but not for flare studies, to the best of our knowledge. {Similarly, TRACE UV filters were also used to deconvolve the \ion{H}{1} Lyman-$\alpha$ channel and the estimate the flare emission of this line \citep{Rubio-da-CostaFletcherLabrosse:2009}.}

The spectral content in the { AIA} UV images has seldom been explored { due to the lack of high-resolution flare spectra in this wavelength range}. The few instruments that have covered this spectral range in the past include the SO82B spectrograph \citep{BartoeBruecknerPurcell:1977} on the Naval Research Laboratory/Apollo Telescope Mount (NRL/ATM) on {\em Skylab}, the {\em SOLar STellar Irradiance Comparison Experiment} (SOLSTICE) \citep{BrekkeRottmanFontenla:1996}, and  {\em Solar Ultraviolet Measurements of Emitted Radiation} (SUMER), on the {\em Solar and Heliospheric Observatory} (SOHO) mission \citep{GontikakisWinebargerPatsourakos:2013}. In 1973--1974, the SO82B spectrograph made observations of a few flares that provided the means for remarkable advances in our understanding of the flaring atmosphere, the formation of spectral lines and continua, emission measure distribution, electron densities, and mass motions via Doppler-shift analysis \citep[e.g.][]{DoschekVanhoosierBartoe:1976,FeldmanDoschekRosenberg:1977,Kjeldseth-MoeNicolas:1977,DereHoranKreplin:1977,CanfieldCook:1978,Withbroe:1978,ChengKjeldseth-Moe:1978,Cheng:1978,LitesCook:1979,CookBrueckner:1979,DereCook:1979,DoyleWiding:1990,WidingDoyle:1990}. This knowledge of the UV flare spectrum has not been fully applied to the analysis of more recent UV imaging.  
The {\em Interface Region Imaging Spectrograph} \citep[{\em IRIS};][]{De-PontieuTitleLemen:2014} covers the UV ranges 1332--1407~\AA\ and 2783--2837~\AA, with high spatial and spectral resolution. While IRIS has allowed many advances in our comprehension of flare physics \citep[e.g.][]{GrahamCauzzi:2015, KerrSimoesQiu:2015},  
there is no overlap with the wavelength range covered by AIA. { However, it is worth noting that efforts have been made to identify the spectral content in the IRIS broadband 2832~\AA\ slit-jaw images (SJI). \cite{KleintHeinzelKrucker:2017} convolved the observed IRIS flare spectra with the SJI 2832~\AA\ response function and found that spectral lines have a significant contribution to the images, along with a possible contribution from the \ion{H}{0} Balmer continuum.
\cite{KowalskiAllredDaw:2017} also identified that many \ion{Fe}{2} lines account for a significant contribution during flares and their models suggest that excess intensity in SJI 2832~\AA\ have similar contributions from \ion{Fe}{2} lines and continuum emission. 
}

In this paper, we use high-resolution spectra in the UV range 1400--1960~\AA\ provided by the Skylab NRL SO82B spectrograph to estimate the spectral contributions to the AIA UV passbands, both during flares { and in quiescent plage regions. Definitive identification of the dominant spectral features, and their respective formation temperatures, will provide valuable  knowledge on where in the solar atmosphere the observed emission is formed.} 
This will { potentially allow} plasma diagnostic information to be obtained from these two filters { in the future}. The calibration method of the SO82B data is described in the Appendix \ref{ap:skylab}.

\section{Skylab NRL SO82B spectrograph}
\label{sec:overview}

We employ data from the SO82B spectroheliograph on Skylab NRL's Apollo Telescope Mount Experiment. The data and supporting documents are available online\footnote{\url{http://louis14.nrl.navy.mil/skylab/}}. 

The spectrograph observed through a slit of 2$''$$\times$60$''$ (1450$\times$43500~km), with no spatial resolution along the slit. The instrument covered the wavelength range 970--1960~\AA with a spectral resolution of 0.06~\AA. SO82B provided UV spectra of solar flares with perhaps the highest spectral resolution, available over a wide spectral range, and with sufficiently small field-of-view so that contributions from quiescent regions did not affect the observed spectra. 
We refer the reader to \cite{CookBrueckner:1979}, \cite{DoyleCook:1992}, and references therein for more details and references about the instrument and data calibration. 

\subsection{Overview of SOL1973-09-07 X1} \label{sec:obs}
\begin{table*}[!t]
\begin{center}
\caption{Skylab NRL SO82B Observations \label{tab:info}}
   \begin{tabular}{llllll}
\hline
\hline  
Source &Date	&Time	&Active Region	&Exposure time	&Film strip \\
\hline 
Flare & 1973-Sept-07 & 12:12~UT &NOAA AR209 & 20.25~s & \verb!nrl_2B450_004_v01! \\
Plage & 1973-Sept-11 & 02:20~UT &NOAA AR219 & 39.00~s & \verb!nrl_2B476_007_v01! \\
\hline
   \end{tabular}
\end{center}
\end{table*}

\begin{figure*}[!t]
\resizebox{\hsize}{!}{\includegraphics[angle=0,trim=0 0 20 0,clip]{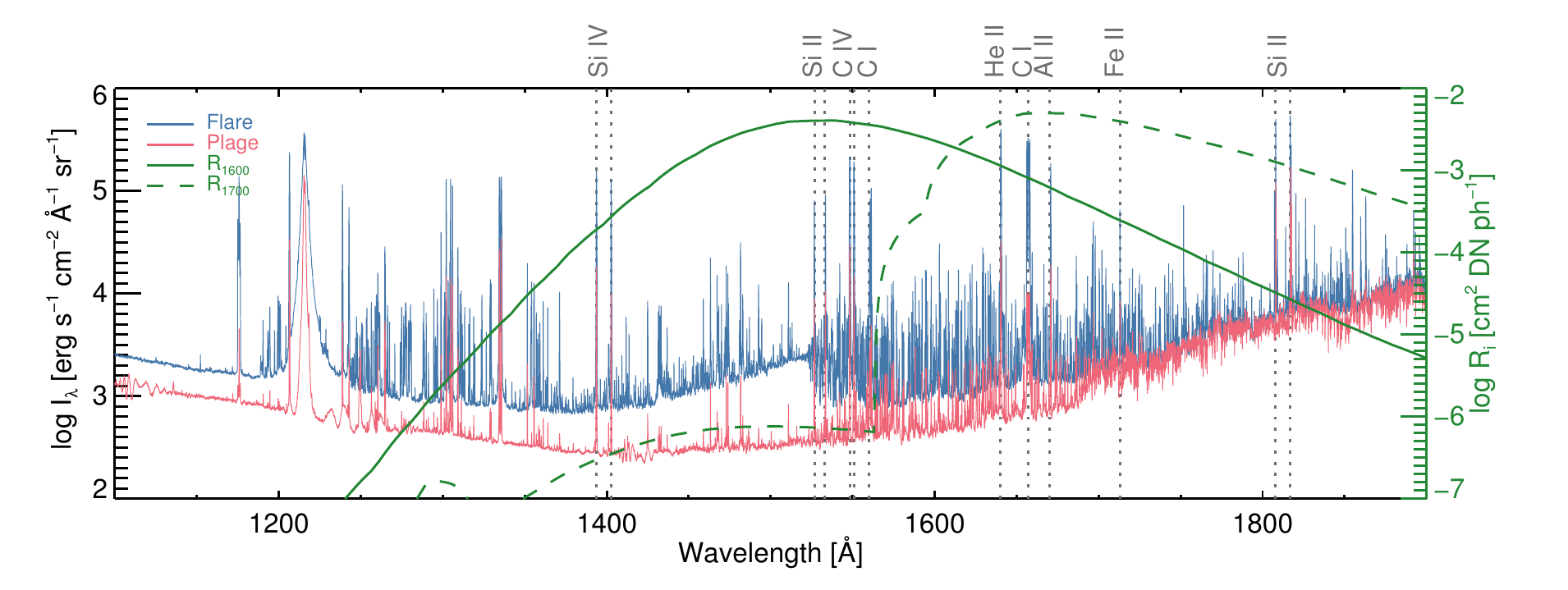}}
\caption{SO82B spectra of the flare SOL1973-09-07 (blue) and plage (red) and response functions, $R_i$, of the AIA UV filters 1600~\AA\ and 1700~\AA. The stronger lines in the spectral range of AIA are indicated in the figure.}
\label{fig:spectra}
\end{figure*}

The solar flare SOL1973-09-07, an X-ray class X1 event (classified by {\em Solrad 9} observations in the 1--8~\AA\ band), occurred in active region NOAA AR209, located at S22W43 { (with a heliocentric cosine angle, $\mu=0.63$)}. SO82B exposures began 12 minutes after the soft X-ray peak, { which corresponded to} the gradual phase of the flare. The event was a two-ribbon flare, as observed in H$\alpha$ images. According to \cite{CookBrueckner:1979}, the spectrograph slit was entirely positioned within the flaring area, therefore no filling factor corrections are necessary. The spectrum of a plage region in active region NOAA AR219 observed on 1973 September 11 ($\mu=0.74$), { just four days after the flare}, is used as a reference for the quiet Sun. The information regarding the flare and plage observations are described in Table~\ref{tab:info}. Further details and analysis of the flare can be found elsewhere \citep{CookBrueckner:1979,DoyleWiding:1990,DoylePhillips:1992}. 

The calibrated spectra\footnote{The calibrated spectra are available as CSV files at \url{http://dx.doi.org/10.5525/gla.researchdata.681}} for the flare SOL1973-09-07 and for the plage are presented in Figure~\ref{fig:spectra} \citep[see also][]{DoyleCook:1992}, along with the response functions of the AIA UV filters 1600~\AA\ and 1700~\AA. Our data calibration method is described in Appendix~\ref{ap:skylab}.

From Figure~\ref{fig:spectra}, it is clear that most, if not all, spectral lines in this wavelength range are enhanced during the flare. Here we focus on the lines within the AIA UV range, $1400 < \lambda < 1850$ \AA. The most prominent lines in this range are \ion{Si}{2} 1526.7 and 1533.4~\AA, \ion{C}{4} 1548.2 and 1550.8~\AA, \ion{C}{1} multiplets around 1560.7 and 1656.3~\AA, \ion{He}{2} 1640.4~\AA, \ion{Al}{2} 1670.8~\AA, \ion{Fe}{2} 1713.1~\AA, and \ion{Si}{2} 1808.0 and 1816.9~\AA, all of which are indicated in Figure~\ref{fig:spectra}. Many other lines, mostly from neutral and weakly-ionized metals (e.g. \ion{C}{1}, \ion{Si}{1}, \ion{S}{1}, \ion{O}{1}, \ion{Fe}{2}) are also enhanced during the flare. There is also a clear enhancement of different parts of the continuum, especially the \ion{Si}{1} $3p^2$ $^3$P {continuum with a recombination edge at} 1525~\AA. The continuum at $\lambda > 1700$~\AA\ does not show any {flare-related} increase, similar to other events observed in this range \citep{CookBrueckner:1979,ChengKjeldseth-Moe:1978,BrekkeRottmanFontenla:1996}. The measured intensity values (in erg~s$^{-1}$~cm$^{-2}$~sr$^{-1}$) of the strongest lines and continua (further discussed in Section~\ref{sec:cont}) were obtained by integrating the specific intensity spectrum, $I(\lambda)$, of each spectral feature in both flare and plage spectra: 
\begin{equation}
I = \int_{\lambda_1}^{\lambda_2} I_\lambda d\lambda.
\label{eq:integral}
\end{equation}
The underlying continuum was subtracted to calculate the line intensity values. The results are presented in the upper part of Table~\ref{tab:lines}. Hereafter, the lines indicated in this Table will be referred to as {\em strong lines}, while the weaker lines are referred to as {\em other lines}.

\begin{table}
\begin{center}
\caption{Main spectral features in the 1400--1900~\AA\ range. \label{tab:lines}}
   \begin{tabular}{llcccc}
\hline
\hline 
Ion & $\lambda$ (\AA) & $\log T$ (K) & \multicolumn{3}{c}{I (10$^5$~erg~s$^{-1}$~cm$^{-2}$ sr$^{-1}$)} \\
&  &  & Flare & Plage & Flare excess \\
\hline
\hline
\ion{Si}{2} & 1526.7 & 4.5 & 0.16 & 0.02 & 0.15                   \\ 
\ion{Si}{2} & 1533.4 & 4.5 & 0.21 & 0.02 & 0.19                   \\ 
\ion{C}{4} & 1548.2 & 5.0 & 1.14 & 0.08 & 1.06                    \\ 
\ion{C}{4} & 1550.8 & 5.0 & 0.87 & 0.04 & 0.83                    \\ 
\ion{C}{1}\tablenotemark{a} & 1560.7 & 4.2 & 0.59 & 0.03 & 0.57  \\ 
\ion{He}{2} & 1640.4 & 4.9 & 2.08 & 0.10 & 1.98                   \\ 
\ion{C}{1}\tablenotemark{b} & 1656.3 & 4.2 & 3.64 & 0.10 & 3.53  \\ 
\ion{Al}{2} & 1670.8 & 4.5 & 0.52 & 0.04 & 0.48                   \\ 
\ion{Fe}{2} & 1713.1 & 4.5 & 0.11 & 0.02 & 0.09                   \\ 
\ion{Si}{2}\tablenotemark{c} & 1808.0 & 4.5 & 1.14 & 0.20 & 0.94  \\ 
\ion{Si}{2}\tablenotemark{c} & 1816.9 & 4.5 & 1.52 & 0.39 & 1.12  \\ 
 \hline
\multicolumn{3}{l}{cont. $1430<\lambda<1525$~\AA} & 1.53 & 0.31 & 1.22 \\ 
\multicolumn{3}{l}{cont. $1525 < \lambda < 1700$~\AA} & 2.07 & 1.01 & 1.05 \\ 
\multicolumn{3}{l}{cont. $1700 < \lambda < 1960$~\AA} & 19.61 & 19.61 & 0.00 \\ 
\hline

\multicolumn{3}{l}{Total (strong lines) } & 11.98 & 1.04 & 10.94 \\ 
\multicolumn{3}{l}{Total (other lines) } & 11.21 & 0.23 & 10.98 \\ 
\multicolumn{3}{l}{Total (cont) } & 23.31 & 20.93 & 2.28 \\ 
\hline
\multicolumn{3}{l}{Total (cont+lines)} & 46.40 & 22.20 & 24.20 \\ 
\hline
\end{tabular}
\tablenotetext{a}{triplet of \ion{C}{1} 1561.3, 1560.7 and 1561.4~\AA.}
\tablenotetext{b}{blend of \ion{C}{1} 1656.3, 1656.9, 1657.0, 1657.4, 1657.9, 1658.1~\AA. and \ion{Fe}{2} 1656.7~\AA.}
\tablenotetext{c}{blend/contribution of \ion{Si}{1}.}
\end{center}
\end{table}

\begin{figure*}
\resizebox{\hsize}{!}{\includegraphics[angle=0,trim=0 0 40 0,clip]{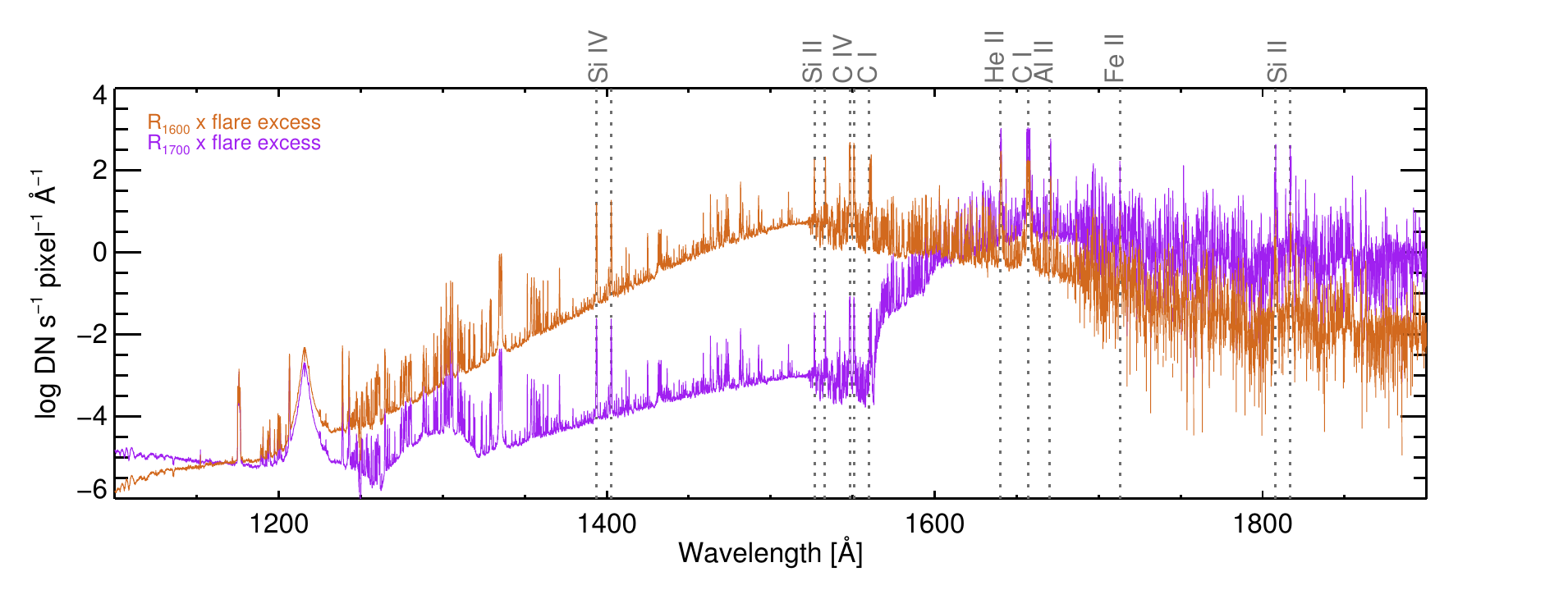}}
\caption{SO82B flare excess spectrum of SOL1973-09-07 convolved with the response functions $R_i$ of AIA 1600~\AA\ (brown) and 1700~\AA\ (purple). The stronger lines in the spectral range of AIA are indicated in the figure.}
\label{fig:specconv}
\end{figure*}

\begin{figure*}
\resizebox{\hsize}{!}{\includegraphics[angle=0]{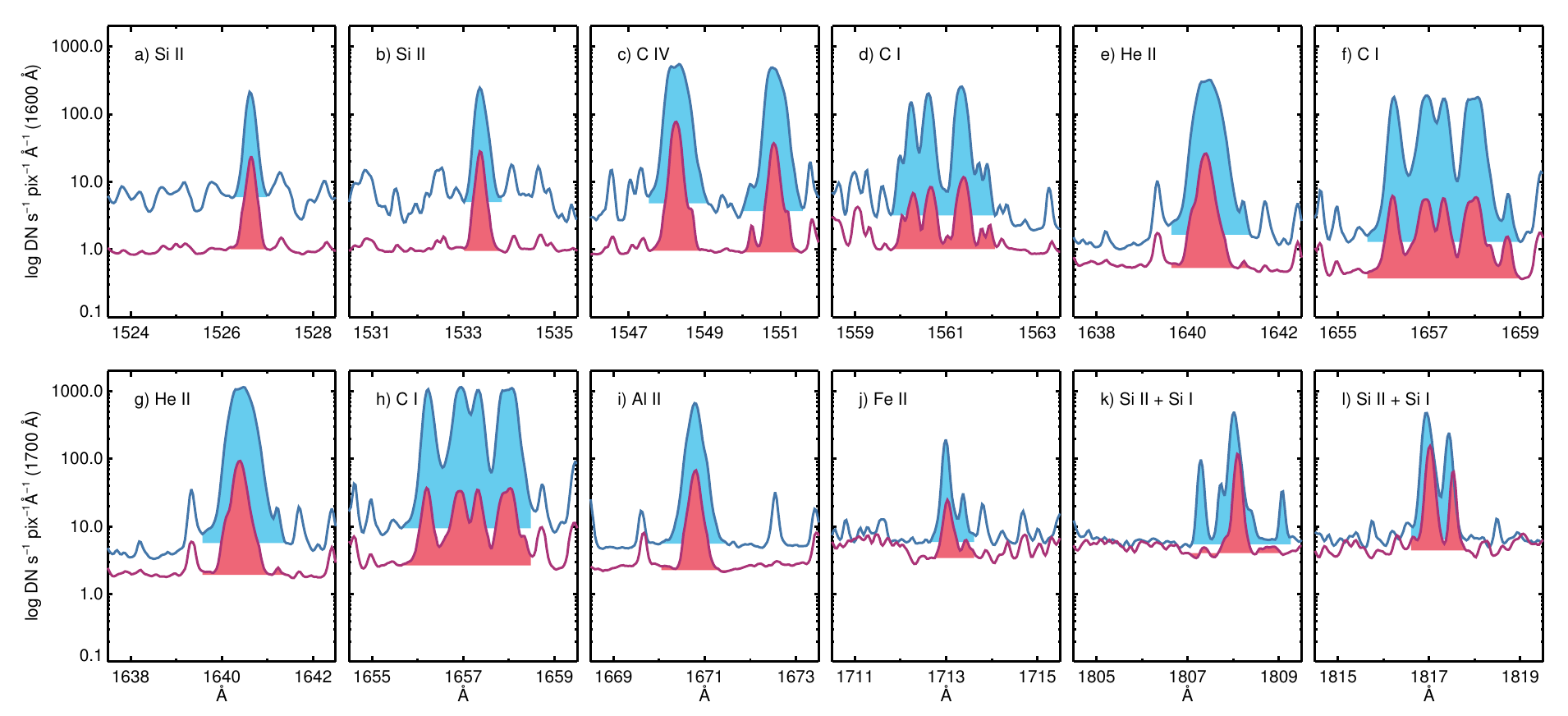}}
\caption{Detailed spectra of { most of the} strongest spectral lines in the SO82B spectra of the SOL1973-09-07 flare (blue) and plage observed on 11 September 1973 (red) convolved with the response functions $R_i$ of AIA 1600~\AA\ (top row) and 1700~\AA\ (bottom row). The shaded areas represent the wavelength range for integration to obtain the line intensity.} 
\label{fig:lineconv}
\end{figure*}

\section{Contributions of spectral features to AIA UV images}

\subsection{Response of AIA UV filters}

AIA records images of the entire Sun with a spatial resolution of about 1.3$''$, using 9 filters: seven EUV channels (94, 131, 171, 193, 211, 304 and 335~\AA) and two UV channels (1600 and 1700~\AA). The cadence of the images is 12 seconds for the EUV channels, and 24 seconds for the UV channels. Each pixel in an image will have a {count or `data number' ($\mathrm{DN}$)} value for wavelength channel, $i$, given by the integrated specific intensity, $I(\lambda, \Omega)$, over the solid angle $\Omega$ subtended by the pixel:
\begin{equation}
\mathrm{DN} = \int_0^\infty \int_\mathrm{pixel} R_i(\lambda) I(\lambda,\Omega) d\Omega d\lambda
\end{equation}
where $R_i$ is the wavelength response function of channel $i$ of the telescope. 

The responses of the UV filters on AIA were obtained via the standard SolarSoftWare \citep[SSW;][]{FreelandHandy:1998} routine \verb!aia_get_response! \citep{BoernerEdwardsLemen:2012}, with keywords {\em area} {\em uv}, {\em dn} set. The response function of each channel, $R_i$, is the product of the effective area, $A_\mathrm{eff}(\lambda)$, (in cm$^2$) and the gain, $G(\lambda)$, (DN photon$^{-1}$). Degradation over time of the UV channels has been identified by the AIA instrument team (W.~Liu and M.~Cheung, private comm.) and corrected via cross-calibration with the {\em Solar EUV Experiment} (SEE) instrument on the {\em Thermosphere Ionosphere Mesosphere Energetics and Dynamics} (TIMED) mission \citep{WoodsEparvierBailey:2005}. In the present work we use the original UV filter responses and note that the proper time-dependent correction must be applied to the AIA UV data for any quantitative analysis.

In Figure~\ref{fig:specconv} we present the convolution of the flare excess spectrum with the AIA response functions, $R_i$, of the UV filters 1600~\AA\ and 1700~\AA. The contributions of the spectral lines and continua to the AIA passbands were estimated using Equation~\ref{eq:integral}. We convolved each observed spectrum, $I(\lambda)$, through the response functions, $R_i(\lambda)$, (for each channel $i$, 1600~\AA\ and 1700~\AA) obtaining the equivalent intensity, DN$_i(\lambda)$ (DN s$^{-1}$ pixel$^{-1}$ \AA$^{-1}$). Then, for each spectral feature in both flare and plage spectra we integrated DN$_i(\lambda)$ in wavelength using a simple trapezoidal rule. The flare excess contribution was then found by subtracting the plage intensity from the flare intensity. The results are presented in Table~\ref{tab:DN}. We calculated the total DN values by integrating the entire convolved spectrum, DN(total), which can be directly associated with the DN pixel values in AIA images. These values were used to obtain the contribution from the multitude of {\em other} spectral lines: DN(other lines) = DN(total) $-$ DN(strong lines) $-$ DN(continuum), { presented in Table~\ref{tab:frac}}.

\begin{table*}
\begin{center}
\caption{Contributions of spectral features to the AIA UV filters. \label{tab:DN}}
   \begin{tabular}{lcc rrr rrr}
\hline
\hline 
Ion & $\lambda$  & $\log T$ 
& \multicolumn{3}{c}{AIA 1600~\AA}
& \multicolumn{3}{c}{AIA 1700~\AA}  \\

& (\AA) &  (K) 
& \multicolumn{3}{c}{(DN s$^{-1}$ pix$^{-1}$)} & \multicolumn{3}{c}{(DN s$^{-1}$ pix$^{-1}$)} \\

& & & 
Flare & Plage & Excess
& Flare & Plage & Excess \\

\hline

\ion{Si}{2} & 1526.7 & 4.5   
& 42.8 & 4.4  & 38.4  & - & - & - \\

\ion{Si}{2} & 1533.4 & 4.5 
& 54.9 & 5.3 & 49.6 & - & - & -  \\ 

\ion{C}{4} & 1548.2 & 5.0 
& 292.6 & 21.6 & 271.0 & - & - & -\\ 

\ion{C}{4} & 1550.8 & 5.0 
& 214.6 & 9.3 & 205.3 & - & - & - \\ 

\ion{C}{1} & 1560.7 & 4.2 
& 141.3 & 6.1 & 135.2 & - & - & - \\

\ion{He}{2} & 1640.4 & 4.9 
& 165.7 & 8.0 & 157.7 & 591.2 & 28.4 & 562.8\\

\ion{C}{1} & 1656.3 & 4.2 
& 212.1 & 6.5 & 205.6 & 1274.0 & 36.5 & 1237.5 \\

\ion{Al}{2} & 1670.8 & 4.5 
&   22.9 &    2.0 &   20.9 &  186.3 &   16.0 &  170.4 \\

\ion{Fe}{2} & 1713.1 & 4.5 
&    2.0 &    0.3 &    1.7 &   31.6 &    4.5 &   27.1 \\

\ion{Si}{2} & 1808.0 & 4.5 
&    2.9 &    0.5 &    2.4 &  111.4 &   19.2 &   92.2 \\

\ion{Si}{2} & 1816.9 & 4.5 
&    3.2 &    0.8 &    2.3 &  131.5 &   34.1 &   97.4 \\

\hline
\multicolumn{3}{l}{cont. $1430 < \lambda < 1525$~\AA} 
& 280.6 & 51.8 & 228.8 & - & - & - \\

\multicolumn{3}{l}{cont. $1525 < \lambda < 1700$~\AA} 
& 252.7 & 113.7 & 139.0 & 393.6 & 211.1 & 182.5 \\

\multicolumn{3}{l}{cont. $1700 < \lambda < 1960$~\AA} 
& 34.9 & 34.9 & 0.0 & 1113.3 & 1113.3 & 0.0 \\

\hline
\multicolumn{3}{l}{Total (strong lines) } 
& 1155.0 & 64.8 & 1090.2   & 2326 & 138.7 & 2187.3 \\

\multicolumn{3}{l}{Total (other lines) } 
& 427.4 & 34.5 & 392.9   & 997.4 & 67.7 & 929.7 \\

\multicolumn{3}{l}{Total (cont) } 
& 493.1 & 174.6 & 318.5 & 1331.9 & 1172.0 & 159.9 \\

\multicolumn{3}{l}{Total (cont+lines)} 
& 2150.6 & 299.8 & 1850.8 & 4830.3 & 1530.8 & 3299.5 \\

\hline
\end{tabular}
\end{center}
\end{table*}

\begin{table*}
\begin{center}
\caption{Fractional contributions of spectral features to the AIA UV filters. \label{tab:frac}}
   \begin{tabular}{lcc rrr rrr}
\hline
\hline
Ion & $\lambda$ (\AA) & $\log T (K)$ 
& \multicolumn{3}{c}{AIA 1600~\AA}
& \multicolumn{3}{c}{AIA 1700~\AA}  \\
&   & 
& Flare & Plage & Excess
& Flare & Plage & Excess \\

\hline

\ion{Si}{2} & 1526.7 & 4.5 
& 0.02 & 0.01 &  0.02 & - & - & -  
\\

\ion{Si}{2} & 1533.4 & 4.5 
& 0.03  &  0.02  &  0.03 &   -  &   -   &  -  
\\ 

\ion{C}{4} & 1548.2 & 5.0 
& 0.14  &  0.07  &  0.15 &   -   &  -  &   -  
\\ 

\ion{C}{4} & 1550.8 & 5.0 
&   0.10  &  0.03 &   0.11 &   -   &  -  &   -  
\\ 

\ion{C}{1} & 1560.7 & 4.2 
& 0.07  &  0.02 &   0.07  &  -  &   - &    -   
\\

\ion{He}{2} & 1640.4 & 4.9 
&  0.08  &  0.03  &  0.09  &  0.12 &   0.02  &  0.17 
\\

\ion{C}{1} & 1656.3 & 4.2 
& 0.10  &  0.02 &   0.11 &   0.26   & 0.02 &   0.38 
\\

\ion{Al}{2} & 1670.8 & 4.5 
& 0.01  &  0.01  &  0.01  &  0.04 &   0.01  &  0.05 
\\

\ion{Fe}{2} & 1713.1 & 4.5 
&   -  &   -   &  -  &   0.01  &  $<$0.01 &   0.01 
\\

\ion{Si}{2} & 1808.0 & 4.5 
&  -   &  -  &   -   &  0.02  &  0.01  &  0.03 
\\

\ion{Si}{2} & 1816.9 & 4.5 
& -  &   -  &   -   &  0.03  &  0.02  &  0.03 
\\

\hline
\multicolumn{3}{l}{cont. $1430 < \lambda < 1525$~\AA} 
&  0.13 &   0.17  &  0.12 &   -  &   -  &   -   
\\

\multicolumn{3}{l}{cont. $1525 < \lambda < 1700$~\AA} 
& 0.12  &  0.38 &   0.08   & 0.08 &   0.14  &  0.06 
\\

\multicolumn{3}{l}{cont. $1700 < \lambda < 1960$~\AA} 
& 0.02  &  0.12  &  -   &  0.23 &   0.73 &   -   
\\

\hline
\multicolumn{3}{l}{Total (strong lines) } 
& 0.54  &  0.21 &   0.59 &   0.48 &   0.09 &   0.66 
\\

\multicolumn{3}{l}{Total (other lines) } 
&  0.20  &  0.12  &  0.21 &   0.21  &  0.04  &  0.28  
\\

\multicolumn{3}{l}{Total (cont) } 
&  0.26  &  0.67 &   0.20  &  0.31  &  0.87  &  0.06 
\\

\multicolumn{3}{l}{Total (cont+lines)} 
&1.00 &   1.00  &  1.00   & 1.00 &   1.00  &  1.00 
\\

\hline
\end{tabular}
\end{center}
\end{table*}

\subsection{Spectral lines}

The main spectral lines within the 1600~\AA\ passband are \ion{Si}{2} 1526~\AA\ and 1533~\AA, the \ion{C}{4} 1550 doublet, both \ion{C}{1} lines at 1560~\AA\ and 1656~\AA, \ion{He}{2} 1640~\AA, and \ion{Al}{2} 1670~\AA. For the 1700~\AA\ passband, the main spectral lines are \ion{C}{1} 1656~\AA, \ion{He}{2} 1640~\AA, \ion{Al}{2} 1670~\AA, and \ion{Si}{2} 1808~\AA\ and 1817~\AA. It is important to note that, in both filters, a multitude of other lines also contribute. The detailed convolved spectra for the flare and plage are shown in Figure~\ref{fig:lineconv}, where the shaded areas represent the wavelength range for integration to obtain the line intensity (in DN s$^{-1}$~pixel$^{-1}$).

{ The \ion{Si}{4} doublet at 1393.76 and 1402.77 \AA\ show a strong enhancement during the flare, with an intensity excess of $0.52$ and $0.36 \times 10^5$ erg s$^{-1}$ cm$^{-2}$ sr$^{-1}$ respectively. However, their flare excess contribution is only about 6 DN s$^{-1}$ pixel$^{-1}$ for AIA 1600~\AA\ and negligible for AIA 1700~\AA.}

\subsection{Continuum} \label{sec:cont}

The wide range of the UV spectrum covered by the AIA UV filters is populated by a number of free-bound recombination edges of neutral metals \citep{VernazzaAvrettLoeser:1973,VernazzaAvrettLoeser:1976,VernazzaAvrettLoeser:1981}. The shorter wavelength range $1160 < \lambda < 1430$~\AA, containing recombination continua from \ion{C}{1} $2p^2$ $^3$P (edge at 1100~\AA) and \ion{C}{1} $2p^2$ $^1$D (edge at 1239~\AA), is negligible for the AIA UV filters. The range $1430 < \lambda < 1525$~\AA\ corresponds to the free-bound continuum of \ion{Si}{1} $3p^2$ $^3$P (1525~\AA), while $1525 < \lambda < 1700$~\AA\ corresponds to the free-bound continuum of \ion{Si}{1} $3p^2$ $^1$D (1683~\AA), with very weak contributions from free-bound \ion{Mg}{1} $3s^2$ $^1$S$^0$ (1622~\AA) and \ion{Fe}{1} $4s^2$ $^5$F (1574~\AA). Lastly, $1700 < \lambda < 1960$~\AA\ is mostly associated with the free-bound continua of \ion{Fe}{1} $4s^2$ $^5$D (1768~\AA) and \ion{Si}{1} $3p^2$ $^1$S (1986~\AA).

\begin{figure}
\resizebox{\hsize}{!}{\includegraphics[trim=0 0 40 0,clip]{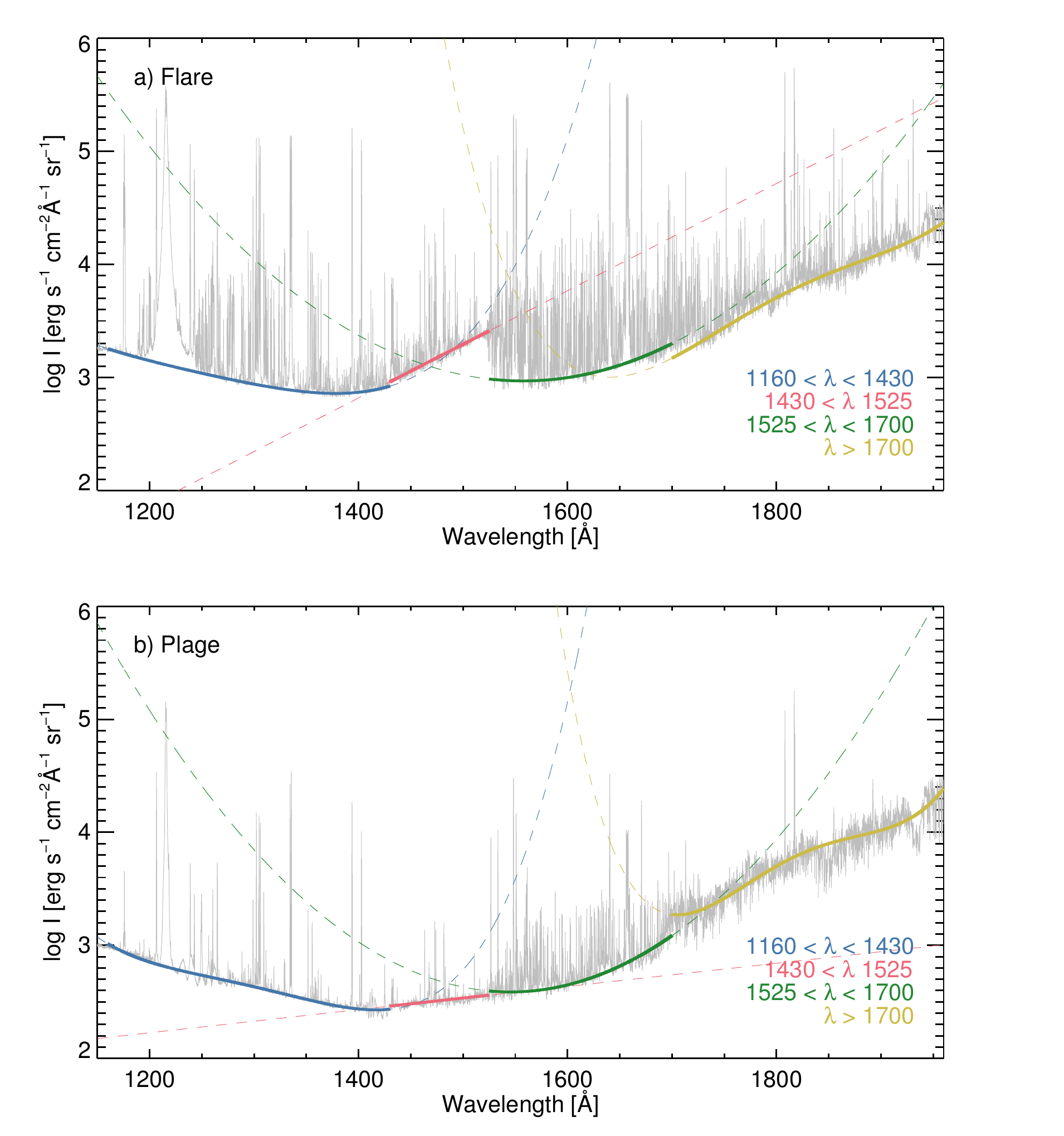}}
\caption{Polynomial fittings for continuum ranges (specified in the legend) for both (a) flare (b) plage spectrum.}
\label{fig:cont1}
\end{figure}

\begin{figure}
\resizebox{\hsize}{!}{\includegraphics[angle=0,trim=0 0 20 0,clip]{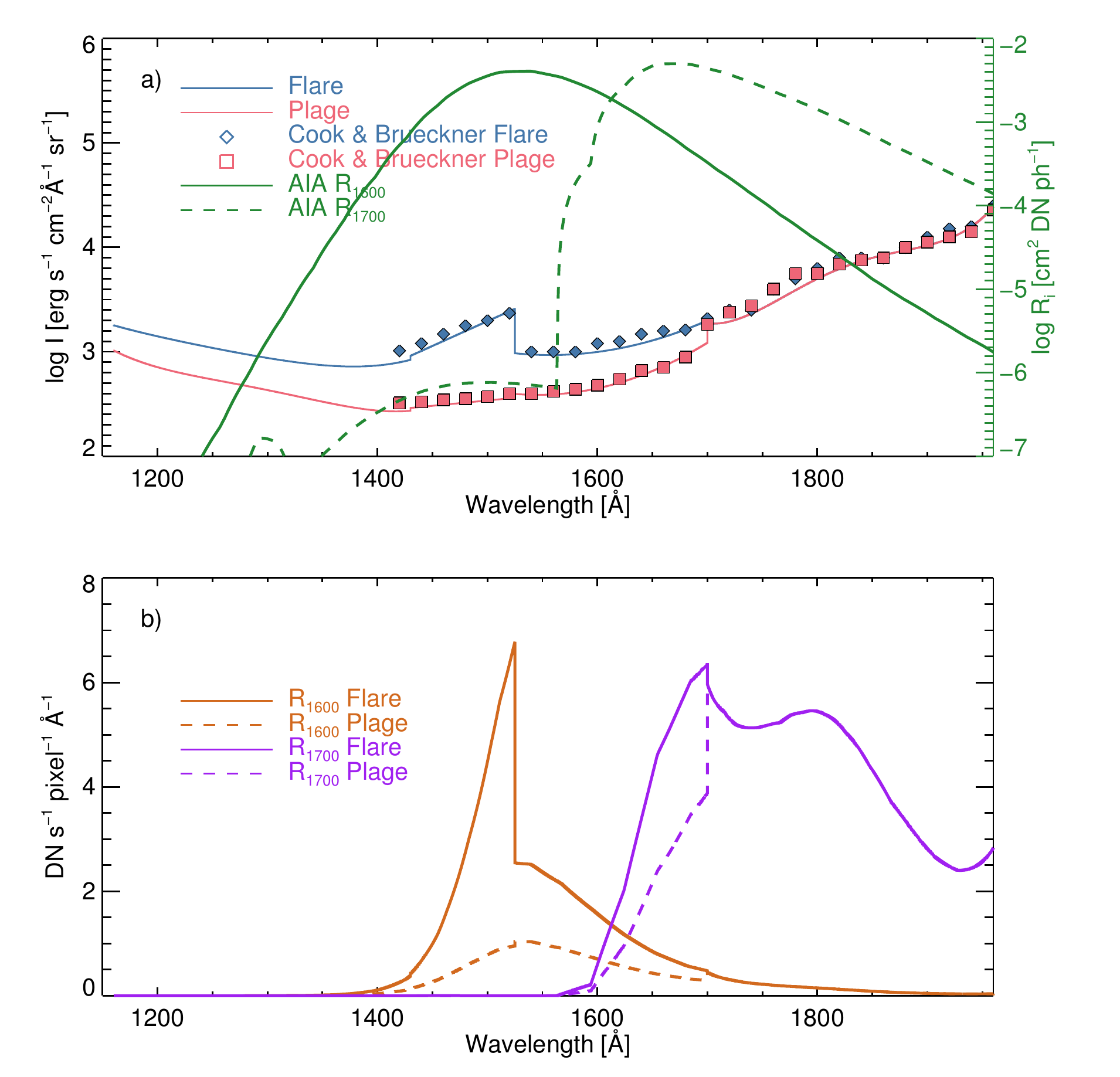}}
\caption{(a) UV continuum of the flare (blue) and plage (red) compared with the continuum values obtained by \cite{CookBrueckner:1979} (blue diamonds and red squares for the flare and plage, respectively). The AIA 1600~\AA\ and 1700~\AA\ response functions are shown for reference. (b) UV continuum convolved with the AIA UV response functions of 1600~\AA\ (brown) and 1700~\AA\ (purple) for both flare (continuous line) and plage (dashed line) spectra. 
}
\label{fig:cont2}
\end{figure}

In order to characterize the continuum, we identified spectral regions without spectral lines, based on the analysis by \cite{CookBrueckner:1979}, and calculated the average specific intensity at those wavelength positions. The resulting values were then fitted with polynomial functions of the wavelength in \AA, of degree $N$,
\begin{equation}
I_\lambda = \sum_{i=0}^{N} a_i \lambda^i
\end{equation}
to represent the continuum at different spectral ranges, $1160 < \lambda < 1430$~\AA\ ($N=4$), $1430 < \lambda < 1525$~\AA\ ($N=1$), $1525 < \lambda < 1700$~\AA\ ($N=2$), and $1700 < \lambda < 1960$~\AA\ ($N=4$), as shown in Figure~\ref{fig:cont1}. The polynomial coefficients are given in Table~\ref{tab:poly}. This method produced a good agreement with the continuum spectrum reported by \cite{CookBrueckner:1979}, for the same observations (Figure~\ref{fig:cont2}a). The recombination edge at 1525~\AA\ of \ion{Si}{1} $3p^2$ $^3$P is evident in the flare spectrum, while no continuum enhancement is detected above 1700~\AA\ (Fig.~\ref{fig:cont1}). The continuum specific intensity, $I_\lambda$, was then integrated in wavelength to obtain the intensity, presented in Table~\ref{tab:lines}. 

The parametric fits that characterize the UV continuum were then convolved them with the AIA response functions, obtaining the DN spectra in Figure~\ref{fig:cont2}b, for both the flare and plage region. These were then integrated in wavelength to obtain the continuum intensity in the same range used for the fitting (Table~\ref{tab:DN}). The continuum enhancement contributing to the AIA 1600~\AA\ filter between $1400 < \lambda < 1700$~\AA, including the \ion{Si}{1} 1525~\AA\ edge, is evident (Fig.~\ref{fig:cont2}b). For the AIA 1700~\AA\ filter, the contribution of the enhanced continuum is { weaker than that of the 1600~\AA\ channel} between $1600 < \lambda < 1700$~\AA.

\begin{table*}
\begin{center}
\caption{Polynomial coefficients for the continuum \label{tab:poly}}
   \begin{tabular}{lccccc}
\multicolumn{6}{c}{Flare coefficients} \\
Range (\AA) & $a_0$ & $a_1$ & $a_2$ & $a_3$ & $a_4$ \\
\hline
\hline
1160--1430 & 4.043e+02 & -1.274e+00 & 1.524e-03 & -8.126e-07 & 1.628e-10 \\
1430--1525 & -3.822e+00 & 4.743e-03 &    -       &    -        &	-   	 \\
1525--1700 & 4.237e+01 & -5.060e-02 & 1.625e-05 &         -   &   -        \\
1700--1960 & 4.482e+03 & -9.936e+00 & 8.247e-03 & -3.037e-06 & 4.187e-10 \\
\hline
\multicolumn{6}{c}{Plage coefficients} \\
Range (\AA) & $a_0$ & $a_1$ & $a_2$ & $a_3$ & $a_4$ \\
\hline
1160--1430 & 1.132e+03 & -3.487e+00 & 4.039e-03 & -2.079e-06 & 4.011e-10 \\
1430--1525 & 9.965e-01 & 1.025e-03  &     -      &     -       &      -     \\
1525--1700 & 5.244e+01 & -6.452e-02 & 2.087e-05 &     -       &      -     \\
1700--1960 & 1.654e+04 & -3.635e+01 & 2.994e-02 & -1.095e-05 & 1.500e-09 \\
\hline
\end{tabular}
\end{center}
\end{table*} 


\subsection{Relative Contributions} \label{sec:rel_cont}

\begin{figure*}
\resizebox{\hsize}{!}{
\includegraphics[trim=100 0 100 0,clip]{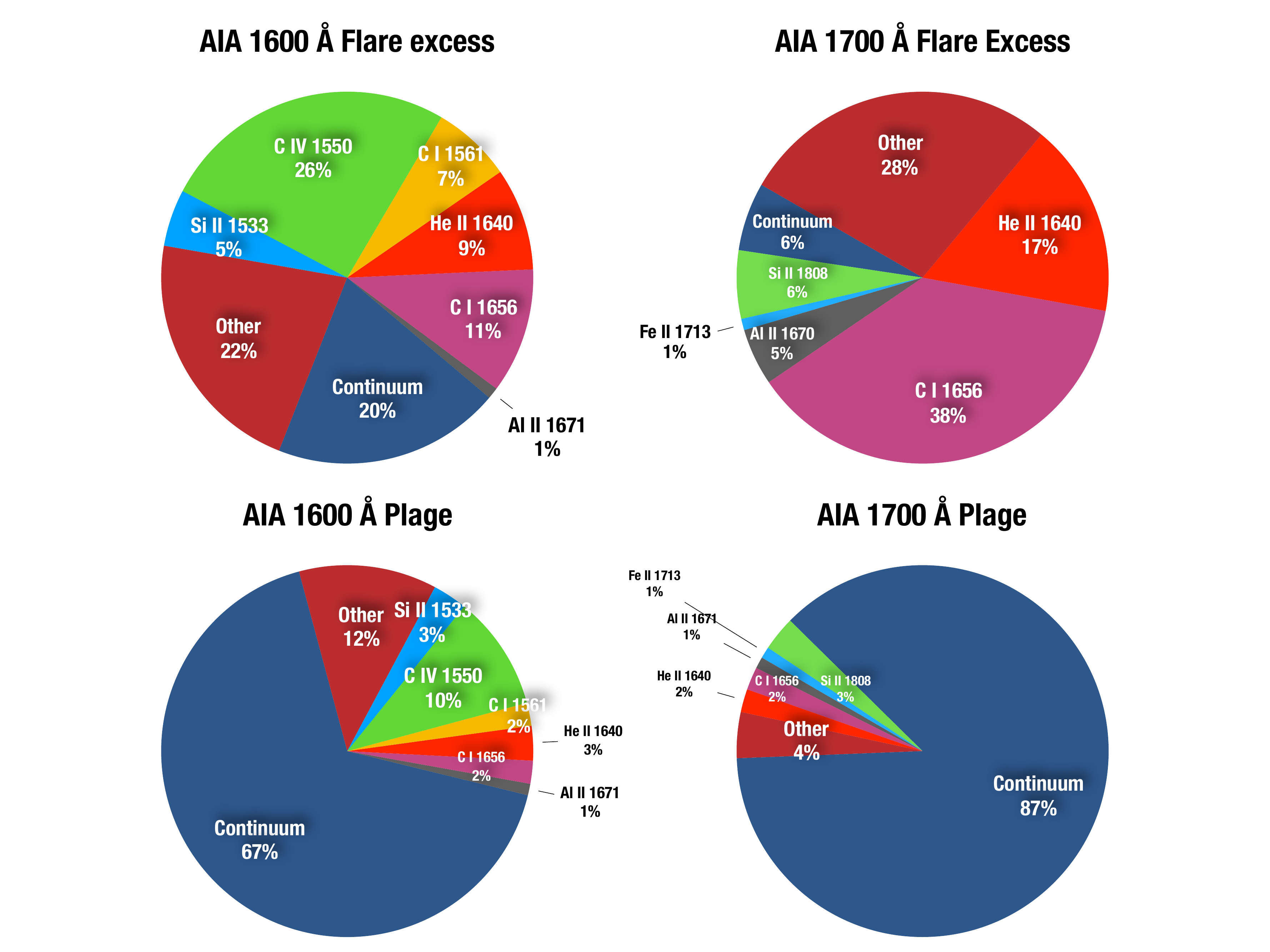}}
\caption{Spectral contributions of the flare excess and plage into AIA 1600~\AA\ and 1700~\AA\ passbands, based on Skylab NRL SO82B data of SOL1973-09-07.}
\label{fig:pie}
\end{figure*}

From the DN values of the main spectral features in the UV spectra in Table~\ref{tab:DN}, we calculated the fractional contributions of each spectral feature $j$ to the AIA UV passbands, defined as DN($j$)/DN(total). All the fractional contributions are shown in Table~\ref{tab:frac} and in Figure~\ref{fig:pie}. The results can be summarized as follows: for AIA 1600~\AA, the strongest {contributors to} the flare excess emission are: 26\% \ion{C}{4} doublet, 20\% continuum (free-bound \ion{Si}{1} $^3$P 1525~\AA\ and \ion{Si}{1} $^1$D 1683~\AA), 31\% from the sum of the \ion{C}{1} multiplets, \ion{He}{2} 1640~\AA, \ion{Si}{2} doublet, and 21\% from the {\em other} lines. For AIA 1700~\AA, the strongest {contributors to} the flare excess emission are: 38\% \ion{C}{1} 1656, 17\% \ion{He}{2} 1640~\AA, 39\% from the {\em other} lines, and only 6\% from continuum. 

For the plage emission, the continuum emission dominates in both filters (Table~\ref{tab:frac} and Figure~\ref{fig:pie}). It amounts to 67\% in 1600~\AA, with 10\% \ion{C}{4}, 11\% from the strong lines and 12\% from the {\em other} lines, and 87\% in 1700~\AA, with 13\% from all spectral lines combined. Different portions of the continuum, formed near the temperature minimum region \citep{VernazzaAvrettLoeser:1981}, contribute in each passband, with $1525 < \lambda < 1700$~\AA\ having the strongest contribution to 1600~\AA\ (\ion{Si}{1} 1683~\AA) and $\lambda > 1700$~\AA\ (\ion{Fe}{1} 1768~\AA\ and \ion{Si}{1} 1986~\AA) to 1700~\AA. This result is not surprising and is, in fact, expected since { quiescent} AIA UV images display clear photospheric characteristics such as network, penumbrae, and umbrae. Evidently, in AIA UV images the flaring pixels contain the emission from both the flaring plasma and underlying atmosphere, and therefore, to compare our flare excess results with the AIA data it would be necessary to subtract the pre-flare emission { from the images}. 

\section{Discussion} \label{sec:discussion}

Our results indicate the dominance of spectral lines { formed} in the temperature range $4.2 < \log T < 5.1$ in the AIA UV flare emission. Although this conclusion is derived from observations during the gradual phase of a flare, the dominance of the spectral lines over the continuum should also occur during the impulsive, { as suggested by the observations of \cite{BrekkeRottmanFontenla:1996}}. The SOLSTICE observations of \cite{BrekkeRottmanFontenla:1996}, taken during the impulsive phase, also show a relatively small continuum enhancement compared to the spectral lines. Therefore, given the temperature range for the formation of the relevant lines, the AIA UV flare excess emission should reflect the response of the chromosphere to the heating and/or energy deposition. This naturally explains the co-temporal and co-spatial hard X-ray and UV emission during flares that has been observed for decades \citep[e.g.][]{WarrenWarshall:2001,AlexanderCoyner:2006,ChengKerrQiu:2012}. Moreover, these results explain the observations of UV bright features at energy release sites near the top of flaring/reconnecting loops as the presence of plasma in the temperature range $4.2 < \log T < 5.1$ \citep{TianLiReeves:2014,SimoesGrahamFletcher:2015,SimoesGrahamFletcher:2015a}. In their analysis of the time evolution of the UV emission from flare ribbons, \cite{ChengKerrQiu:2012} and \cite{QiuSturrockLongcope:2013} assumed that the AIA 1600~\AA\ images are dominated by the \ion{C}{4} 1550~\AA\ doublet plus continuum contribution, and that the AIA 1700~\AA\ images are dominated by continuum emission, ignoring any contribution from other lines. Our results indicate that those assumptions are not entirely valid. However, further analysis of this type using AIA UV images should be encouraged. 

\subsection{Brief comparison of the UV spectrum of solar flares}

Observations of the UV spectrum during flares are rare. Apart from SOL1973-09-07, detected by SO82B, one other observation that is available is that of SOL1992-02-27, presented by \cite{BrekkeRottmanFontenla:1996}. There are some considerable differences in the line intensity values obtained from the SO82B observation of SOL1973-09-07 (Table \ref{tab:lines}) and the estimated $I_f$ values from \citet[see their Table 2]{BrekkeRottmanFontenla:1996}, using the SOLSTICE data of SOL1992-02-27. However, these events were observed at different phases: SOL1992-02-27 was observed during the impulsive phase while SOL1973-09-07 data was taken during its gradual phase. Moreover, the SOLSTICE instrument obtained the spectrum by scanning in wavelength over a period of about four minutes, whereas for SOL1973-09-07 the spectrum was obtained in a single exposure. Therefore, it is not surprising that the line intensities from \cite{BrekkeRottmanFontenla:1996} are larger than the values presented here. 

In the case of in SOL1992-02-27, the \ion{C}{4} 1548.2~\AA\ and 1550.8~\AA\ and \ion{Si}{2} 1526.7~\AA\ and 1533.4~\AA\ doublets were observed near the maximum of the impulsive phase { whereas} lines at longer wavelengths were detected at earlier stages of the flare. { The intensities of the \ion{Si}{2} and \ion{C}{4} lines in SOL1992-02-27 are therefore roughly two orders of magnitude larger than those detected in the gradual phase of SOL1973-09-07 (Table~\ref{tab:lines}). Unfortunately, even with this information it is only possible to affirm that the line intensities are indeed larger, but it is not possible to know their relative intensities, which is the more important aspect in terms of their contribution to impulsive phase emission in the AIA UV filters.}


The differences in the observational methods may also have had an effect, since SO82B observed using a narrow slit positioned over the active region, while SOLSTICE observed the entire Sun without spatial resolution, putting the quiet Sun as the reference background emission. 

Therefore, a direct comparison of the relative line intensities is not possible since they refer to different stages of the flare in SOL1992-02-27. Nevertheless, the same main spectral lines are found in both events. We make this comparison here as a word of caution regarding the interpretation of the spectral components of AIA UV images.


\subsection{Recombination continua in the UV range}

The \ion{Si}{1} continua are often regarded as important contributions to UV images. {Our results indicate clearly that the free-bound continua \ion{Si}{1} $3p^2$ $^3$P 1525~\AA\ and \ion{Si}{1} $3p^2$ $^1$D 1683~\AA\ are the dominant features the AIA 1600~\AA\ images in quiescent times. For AIA 1700~\AA\ images the continua above 1700~\AA, \ion{Fe}{1} $4s^2$ $^5$D (1768~\AA) and \ion{Si}{1} $3p^2$ $^1$S (1986~\AA), are the dominant features.} { \cite{CookBrueckner:1979} present the continuum enhancements of the flare used in our analysis, SOL1973-09-07, and also for SOL1973-08-09, also observed during the gradual phase. The continuum data for both flares are remarkably similar \citep[see Figure 3 in][]{CookBrueckner:1979} which may indicate a general trend for the continuum intensity levels during the gradual phase of flares. Moreover, the observations of SOL1992-02-27 by \cite{BrekkeRottmanFontenla:1996} also show a weak continuum enhancement with respect to the intensity of the spectral lines, further suggesting the relatively small importance of the UV continuum detected by AIA.}

We speculate that the contribution of \ion{Si}{1} and \ion{Fe}{1} continua would be negligible in flaring structures observed off-limb with UV images. As noted in previous works, the \ion{Si}{1} continua are formed around the temperature minimum region, and are enhanced during flares by the radiation field of chromospheric spectral lines \citep{MachadoHenoux:1982,MachadoEmslieMauas:1986,MachadoEmslieAvrett:1989,DoylePhillips:1992}. {Moreover, UV spectroscopic observations by \cite{SamainBonnetGayet:1975} show no off-limb continuum in the range 1460--1750~\AA.} Therefore, flaring structures observed above the limb in UV images should be dominated by chromospheric spectral lines. This should increase the potential use of UV images for many types of studies e.g. in the diagnostic of coronal rain during the late phase of solar flares \citep{AntolinVissersPereira:2015,FromentAuchereMikic:2018}. 

On the other hand, we should also consider the possibility that the flare contribution from \ion{Si}{1} $3p^2$ $^3$P 1525~\AA\ and \ion{Si}{1} $3p^2$ $^1$D 1683~\AA\ continua may occur higher in the atmosphere than in the quiet Sun, possibly as optically thin emission. During flares, the ionization of \ion{H}{0} in the chromosphere increases the electron density to around $10^{13}$--$10^{14}$cm$^{-3}$, as noted in both observations \citep[e.g.][]{Svestka:1972} and radiative hydrodynamic models \citep[e.g.][]{SimoesKerrFletcher:2017}. The excess of free electrons may, under non-equilibrium conditions, force recombination of \ion{Si}{2} into \ion{Si}{1}. This possibility should be investigated using radiative hydrodynamic calculations, such as RADYN \citep{CarlssonStein:1995,AllredKowalskiCarlsson:2015} and Flarix \citep{VaradyKasparovaMoravec:2010,HeinzelKleintKasparova:2017}.
\section{Conclusions}

We have estimated the contributions of UV spectral lines and continua to the 1600~\AA\ and 1700~\AA\ passbands on AIA. Given the lack of current spectroscopic data in the 1300--1900~\AA\ range, we employed observations of the flare SOL1973-09-07 with high spectral resolution made by the NRL SO82B spectrograph on Skylab. We find that the flare excess emission in both AIA UV filters is dominated by spectral lines formed in the chromosphere and transition region, with a temperature range $4.2 < \log T < 5.1$. We also confirm that quiescent plage emission captured by these filters is dominated by the continuum formed in the photosphere.

The calibrated spectrum of the flare and a plage region (used as a reference spectrum during quiescent times) were convolved with the AIA filter responses and the { relative} contributions of the main spectral lines and continua for the flare excess were obtained. We find that in the AIA 1600~\AA\ images the flare excess is composed of 26\% of \ion{C}{4} 1550~\AA, 20\% of \ion{Si}{1} continua, 11\% of \ion{C}{1} 1656~\AA, 9\% of \ion{He}{2} 1640~\AA, 7\% of \ion{C}{1} 1561~\AA, 5\% of \ion{Si}{2} 1526 \& 1533~\AA, plus 21\% from hundreds of weaker lines from neutral and weakly-ionized metals. For AIA 1700~\AA\ images, the flare excess is formed of  38\% \ion{C}{1} 1656~\AA, 17\% \ion{He}{2} 1640~\AA, 12\% from the combined contributions from \ion{Al}{2} 1671~\AA, \ion{Fe}{2} 1713~\AA, \ion{Si}{2} 1808~\AA, and 1817~\AA, 28\% from weaker lines from neutral and weakly-ionized metals, and only 6\% from \ion{Si}{1} $^1$D continuum. The plage emission is dominated by continuum emission: 67\% in the AIA 1600~\AA\ filter and 87\% in the 1700~\AA\ filter. This is expected given the typical photospheric features often seen in the AIA UV images of the quiet Sun.

Our results should be taken as a guideline for the interpretation of the flare emission in the UV bands observed by AIA. These should by no means be taken as the definitive fractions of the flare spectral components in those filters, since flare emission varies dynamically during an event, and also from event to event, so the relative contributions are likely to change. However, these variations should not be extreme, i.e. the \ion{C}{4} 1550~\AA\ doublet and other chromospheric lines should be the main contributors to the flare excess emission detected by AIA 1600~\AA. At the same time, the \ion{C}{1} multiplet at 1656~\AA\ and \ion{He}{2} 1640~\AA\ should account for the largest fraction of the flare excess emission into the AIA 1700~\AA\ passband, with negligible contribution from continua. 

Since spectral observations with high-resolution in this UV range are rare and thus the evolution of the UV spectrum during flares is still largely unknown, radiative hydrodynamic simulations should be used to examine the validity of our results. However, most of the lines in this range are not considered to be energetically important in the solution of the radiative transfer of the flaring plasma and therefore tend to be ignored in the current solutions. The inclusion of \ion{C}{0} and \ion{Si}{0} atomic database in current tools for radiative hydrodynamic modeling should allow us to revisit previous works regarding the behavior of the main spectral lines such as \ion{C}{4} and \ion{C}{1}, and \ion{Si}{1} continua during flares.

Lines of neutral or weakly ionized metals such as \ion{C}{1}, \ion{Si}{1}, \ion{Fe}{2}, \ion{Si}{2}, and \ion{C}{2} are abundant and strongly enhanced during flares \citep{DoyleCook:1992}, and have been useful in the construction of semi-empirical flare atmospheric models \citep[e.g.][]{LitesCook:1979,MachadoAvrettVernazza:1980}. Lines of some of these elements are currently being observed by IRIS \citep[e.g. \ion{Si}{4} 1393.755~\AA and 1402.770~\AA, \ion{C}{2} 1334.535~\AA\ and 1335.708~\AA; see][]{PolitoReepReeves:2016,WarrenReepCrump:2016,ReepWarrenCrump:2016}, and could potentially be used to estimate the intensity of the lines of the same elements at longer wavelengths, within the AIA UV range. 

Armed with this new information about the spectral composition of AIA UV images, joint analysis of IRIS and AIA UV data may help to reveal more about the UV spectrum of flares and quiescent solar features and also to explore the AIA data since 2010 for long-term variability of several solar structures \citep[e.g.][]{Oliveira-e-SilvaSelhorstSimoes:2016}.

\acknowledgements

We would like to express our gratitude to NRL for maintaining an online catalog of the SO82B data archive and its manual. P.J.A.S. and R.O.M. acknowledge support from the University of Glasgow's Lord Kelvin Adam Smith Leadership Fellowship. R.O.M. would also like to acknowledge support from the Science and Technologies Facilities Council for the award of an Ernest Rutherford Fellowship (ST/N004981/1). L.F. and H.A.S.R. acknowledge support from grant ST/P000533/1 made by the UK's Science and Technology Facilities Council. The research leading to these results has received funding from the European Community's Seventh Framework Programme (FP7/2007-2013) under grant agreement no. 606862 (F-CHROMA). Color-blind safe and print-friendly color tables designed by Paul Tol were used in preparing the figures in the manuscript (\url{http://www.sron.nl/~pault/}).

\appendix

\section{Calibration of NRL SO82B film data}
\label{ap:skylab}

Here we describe our method of obtaining the calibrated spectrum of the flare SOL1973-09-07 and plage \citep{DoyleCook:1992} from the digitized film strips. Although film data for other flares do exist, e.g. SOL1973-09-05 \citep{Cheng:1978,ChengKjeldseth-Moe:1978} and SOL1973-08-09 \citep{CookBrueckner:1979}, the values of the specific intensity of spectral lines are not published and prevent the reconstruction of the characteristic curves using our method. We note that specific intensity or line-integrated intensity values (erg~s$^{-1}$~cm$^{-2}$~sr$^{-1}$) of a selection of strong lines have been reported for some other events \citep{Cheng:1978,LitesCook:1979}.

\subsection{Digitized film strips}

\begin{figure*}
\resizebox{\hsize}{!}{\includegraphics[angle=0]{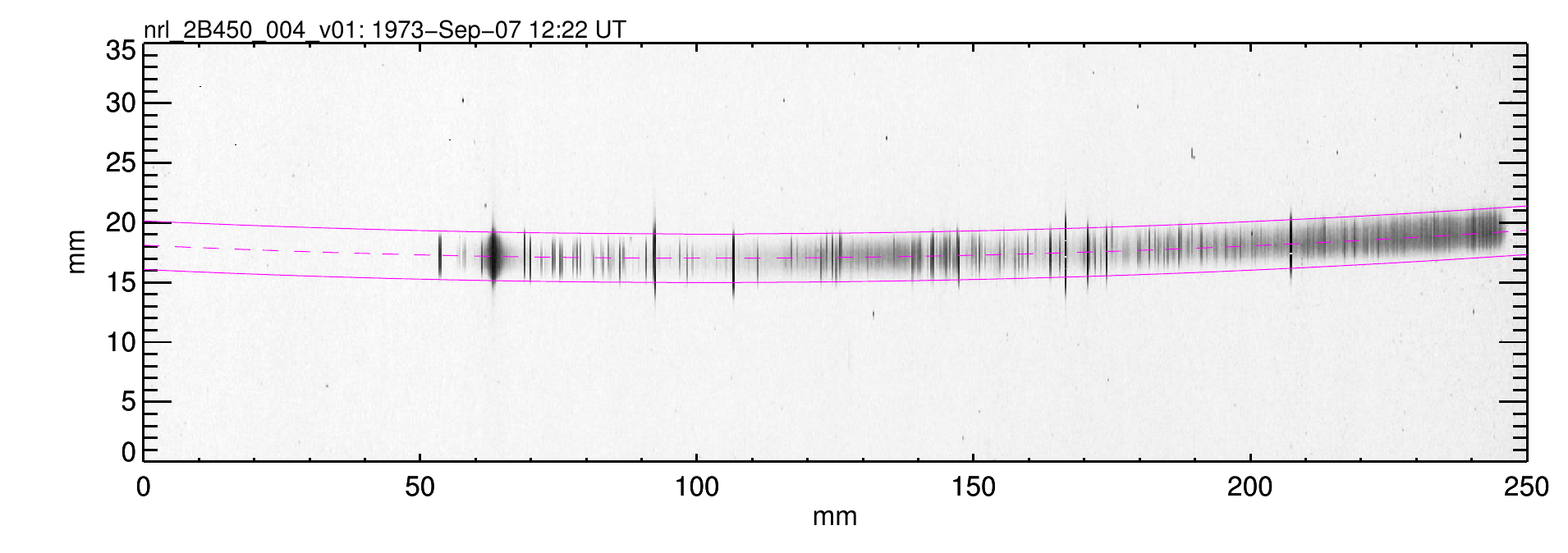}}
\resizebox{\hsize}{!}{\includegraphics[angle=0]{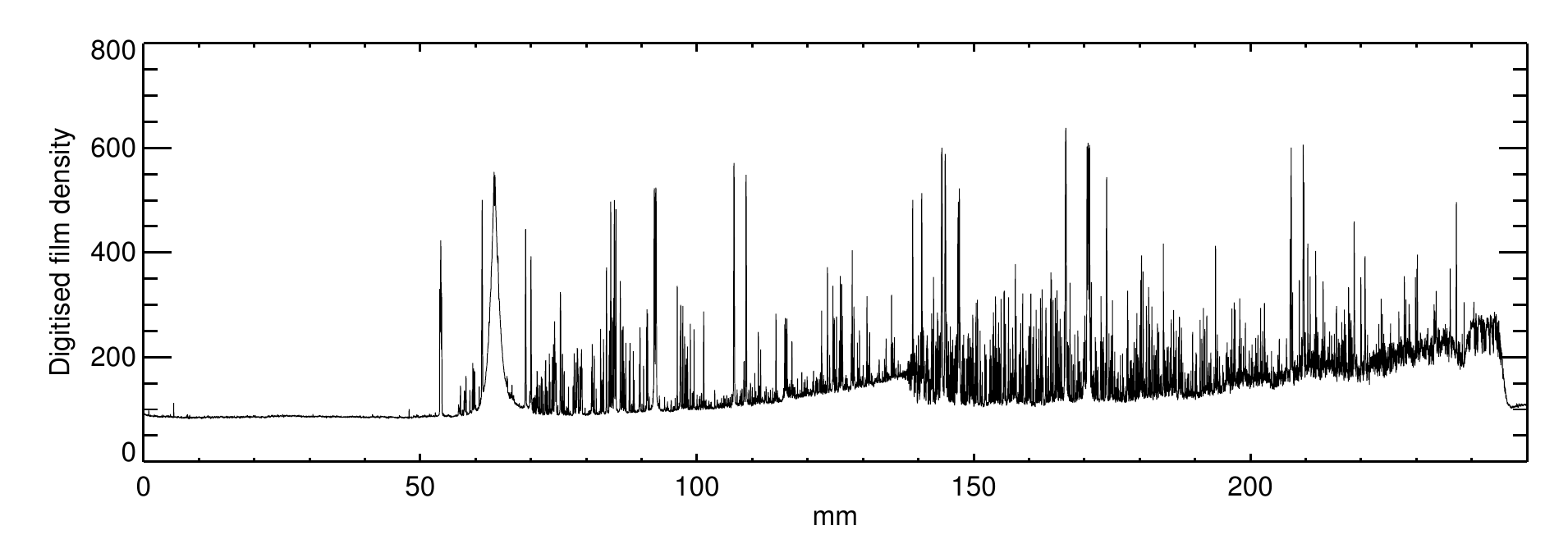}}
\caption{Top: SO82B film plate 2B450-004 for the SOL1973-09-07 flare (not in the correct aspect ratio). The magenta lines indicating the central strip used to acquire the spectrum, shown in the bottom panel, { in raw digital units for the digitized film, before applying the normalization factor $f$.} }
\label{fig:film1}
\end{figure*}

The original SO82B films have a physical size of 35~mm by 250~mm and their high-resolution digitized versions are available as FITS files at \url{http://louis14.nrl.navy.mil/skylab/}, along with the instrument manual, which provides the necessary information to identify the film plate numbers for a given observation. The film plates used in our work are 2B450-004 (flare, shown in the top panel of Figure~\ref{fig:film1}) and 2B476-007 (plage). 

The spectrum registered on the film generally presents a curvature, more pronounced at longer wavelengths. We identified the strongest spectral features which we then used to trace a curve through the center of the registered spectrum. By taking the mean of a band with width of 64 pixels around the central curve we obtain the spectrum in {\em digitized film density} values. The {\em fog} density value (i.e. the film noise), as a function of wavelength position (the long side of the film), was estimated by taking the mean value of digitized film density of two strips above and below the registered spectrum, which was then subtracted from the spectrum. Finally, the spectrum was scaled by a normalization factor $f$, which will be described below. The final digitized film density spectrum for the flare is shown in the bottom panel of Figure~\ref{fig:film1}.

\subsection{Wavelength calibration}
\begin{figure}
\resizebox{\hsize}{!}{\includegraphics[angle=0]{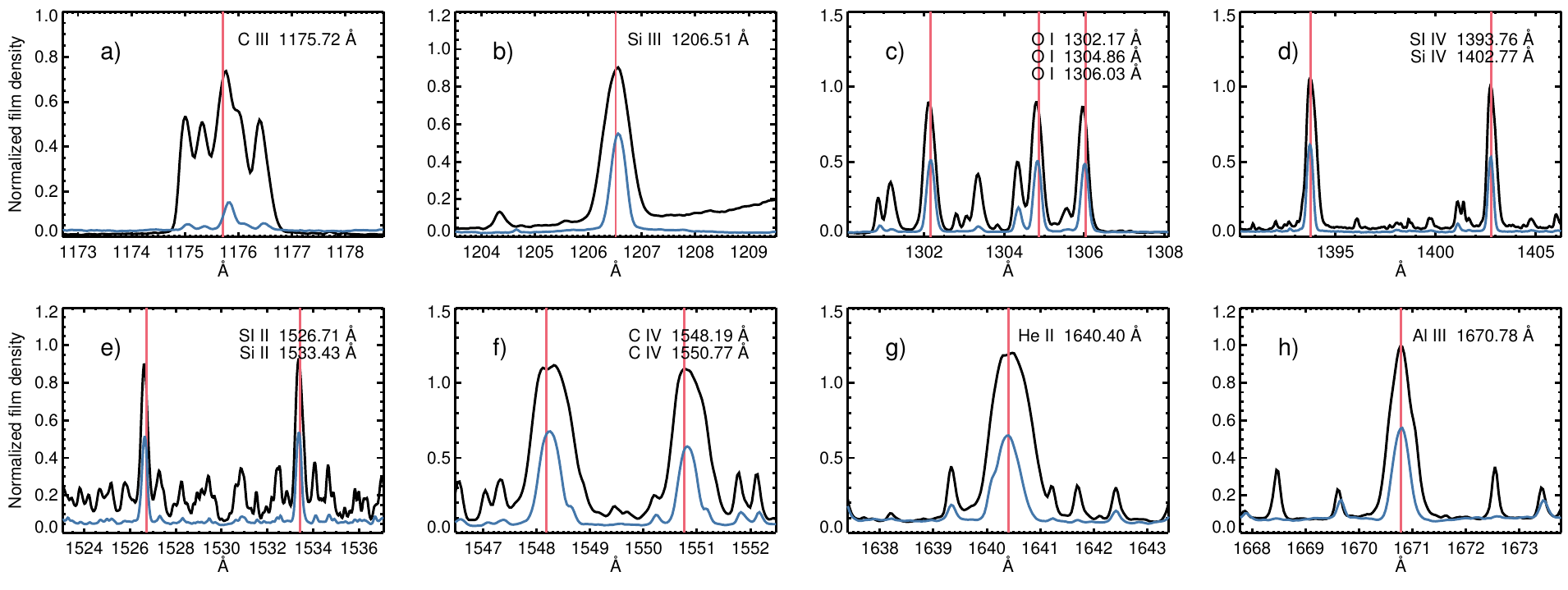}}
\caption{SO82B spectra of the flare (black) and plage (blue) of selected spectral lines used for wavelength calibration, based on wavelengths given by \cite{SandlinBartoeBrueckner:1986} (indicated by the vertical lines). { Note that these spectra were already normalised by the factor $f$ (see text for full explanation).}
}
\label{fig:wavcal}
\end{figure}

The wavelength calibration was done by identifying strong and unblended lines using the references given by \cite{SandlinBartoeBrueckner:1986}. Eventual Doppler shift effects in the lines can be considered small for our purpose (in comparison with the AIA UV filter widths) and can be safely neglected. The selected lines, their reference wavelength and the final wavelength calibration for the flare and plage data are presented in Figure~\ref{fig:wavcal}.
\begin{table}
\begin{center}
\caption{Spectral lines used for wavelength calibration based on the observed wavelengths given by \cite{SandlinBartoeBrueckner:1986}. \label{tab:wav}}
   \begin{tabular}{cc}
Ion & Wavelength (\AA) \\
   \hline
   \hline 
\ion{C}{3} & 1175.72 \\
\ion{Si}{3} & 1206.51 \\
\ion{O}{1} & 1302.17 \\
\ion{O}{1} & 1304.86 \\
\ion{O}{1} & 1306.03 \\
\ion{Si}{4} & 1393.76 \\
\ion{Si}{4} & 1402.77 \\
\ion{Si}{4} & 1526.71 \\
\ion{Si}{4} & 1533.43 \\
\ion{C}{4} & 1548.19 \\
\ion{C}{4} & 1550.77 \\
\ion{He}{2} & 1640.40 \\
\ion{Al}{3} & 1670.78 \\
\hline
   \end{tabular}
\end{center}
\end{table} 

\subsection{Intensity calibration}

The intensity calibration of film data is normally performed by the use of the characteristic curves of the film and calibrations curves of the instrument, with the following expression 
\begin{equation}
\log I_\lambda = \log E_\mathrm{abs} - \log t_\mathrm{exp} + \log E_\mathrm{rel}.
\label{eq:cal}
\end{equation}
{ Here $E_\mathrm{abs}$ is the instrument's exposure indicating the specific intensity $I$ (in units of erg s$^{-1}$ cm$^{-2}$ \AA$^{-1}$ sr${^-1}$) required to produce a {\em film density} of $D=0.3$ in $t=1$ second, and $E_\mathrm{rel}$ is the relative intensity registered by the film. }

\begin{figure}
\centering
\resizebox{0.5\hsize}{!}{\includegraphics[angle=0]{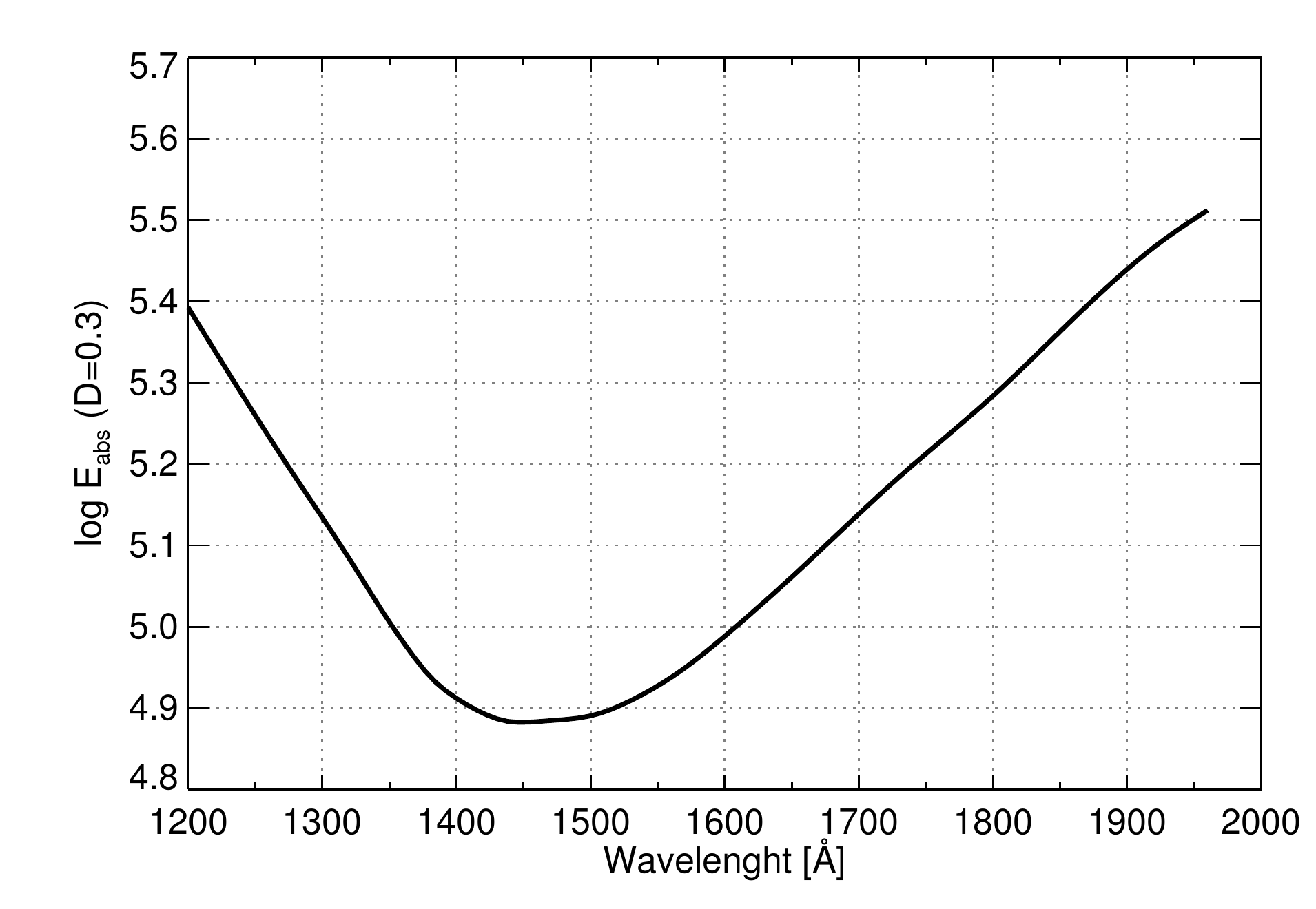}}
\caption{Absolute intensity calibration curve of SO82B, originally published as Figure 5.2 of the spectrograph manual.}
\label{fig:calcurve}
\end{figure}

The term $\log E_\mathrm{abs}$ is obtained from the absolute intensity calibration curve \citep{Nicolas:1977,Kjeldseth-MoeNicolas:1977}, with an overall uncertainty in the range 25--50\% \citep{DoyleCook:1992}, { and it is applicable to any observations made with SO82B}. This curve was originally published as Figure 5.2 of the SO82B manual, reproduced here in Figure~\ref{fig:calcurve}. 

The {\em film exposure} $\log E_\mathrm{rel}$ is obtained from the characteristic curves of the film and the {\em film density} $D$ of the observation. These curves indicate the non-linear association of $\log E_\mathrm{rel}$ and $D$. Finally, $\log t_\mathrm{exp}$ is the time exposure.

According to the SO82B manual, there is no single characteristic curve that is appropriate for all observations. Each film batch has its own non-linear response. \cite{CookEwingSutton:1988} presented a method to obtain the characteristic curves from film strips. Although it might be possible to apply that technique to digitized film data, the method is based on the comparison of films taken with two different exposure times and it relies on the assumption that the spectral lines have the same intensity. Evidently, this method cannot be used for flare data since the intensity of the spectral lines vary during events. We first experimented with the sample curves presented in the SO82B manual and elsewhere \citep{DoschekVanhoosierBartoe:1976}. The utilization of such characteristic curves require film density values, which are not available from the digitized films; we experimented applying arbitrary scaling factors to convert the digitized film values back to analog values. Unfortunately, none of our initial attempts were successful in obtaining specific intensity values as presented by \cite{DoyleCook:1992} or \cite{CookBrueckner:1979} (for the continuum).

The solution was then to re-construct the characteristic curves by comparing the digitized film values with physical values sampled from the published spectra in \cite{DoyleCook:1992}. Using digitizer software\footnote{\url{http://plotdigitizer.sourceforge.net}} we sampled $\log I$ values of several lines and continuum positions from both flare and plage spectra. These values were then converted to $\log E_\mathrm{rel}$ using Equation~\ref{eq:cal}. We verified that the errors in digitizing the data are of the order of 0.016 dex\footnote{$\mathrm{dex}(x)=10^x$}, and are relatively small compared to other uncertainties in the data \citep{DoyleCook:1992}. The time exposures $t_\mathrm{exp}$ for the flare and plage data are given in Table~\ref{tab:info}.

\begin{figure}
\resizebox{\hsize}{!}{\includegraphics[angle=0]{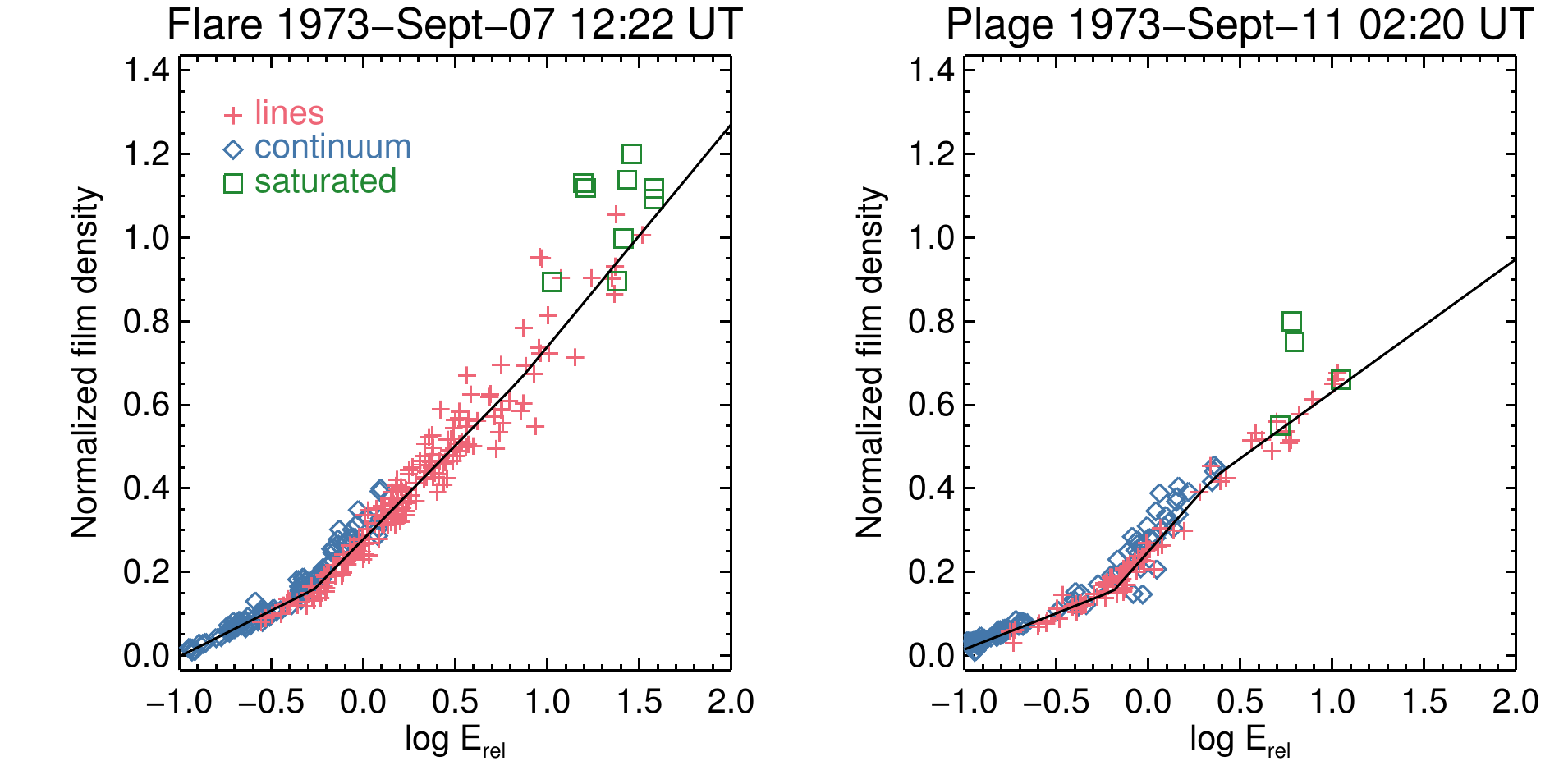}}
\caption{Reconstructed characteristic curves for the (a) flare and (b) plage { normalised data}, using several measured data points in the continuum (blue diamonds) and lines (red crosses). The saturated (green squares) were not used in the fitting. }
\label{fig:hdfinal}
\end{figure}

Values of the digitized film $D$ were then obtained for the same wavelength positions of the spectral points read from \cite{DoyleCook:1992}. The final characteristic curve is then obtained by fitting a function to the points of $D$ (digitized data) as a function of $\log E_\mathrm{rel}$ (reference). We chose a triple broken-line function (Equation~\ref{eq:bline}) 
\begin{eqnarray}
D &=& a_1 x + b_1, x < x_1 \\
D &=& a_2 x + b_2, x_1 < x < x_2 \\
D &=& a_3 x + b_3, x > x_2 
\label{eq:bline}
\end{eqnarray}
where $x=\log E_\mathrm{rel}$, and $b_2 = (a_1-a_2)x_1+b_1$ and $b_3=(a_2-a_3)x_2+b_2$. This function was selected as a way to approximate the linear and non-linear (at film density extremes) properties of the film, and the fitting was performed using the non-linear least-squares fitting routine \verb!mpfitfun! \citep{Markwardt:2009}. The parameters for the reconstructed characteristic curve are given in Table~\ref{tab:par}. The reconstructed characteristic curve, along with the values $\log E_\mathrm{rel}$ and $D$ used, are shown in Figure~\ref{fig:hdfinal}. The red crosses represent points read from spectral lines and blue diamonds from continuum. The green squares represent strong spectral lines for which peak intensities are not available from \cite{DoyleCook:1992} and were not used in the fitting procedure. During the fitting procedure, we iteratively adjusted the normalization factor $f$ in order to make the characteristic curves normalized so that $D=0.3$ refers to $\log E_\mathrm{rel} = 0$, in order that the absolute calibration curve (Figure~\ref{fig:calcurve} normalized in similar way) could be used, { as described in the SO82B manual and in \cite{CookEwingSutton:1988} }. The final values for $f$ are 1.2 and 0.8 for the flare and plage data, respectively. 
\begin{table}
\begin{center}
\caption{Parameters for the reconstructed characteristic curves of SO82B films. 
  \label{tab:par}}
   \begin{tabular}{ccc}
Parameter & Flare & Plage \\
\hline
\hline 
$a_1$ & $0.22\pm0.01$ & $0.17\pm0.01$ \\
$b_1$ & $0.22\pm0.01$ & $0.19\pm0.01$ \\
$a_2$ & $0.45\pm0.02$ & $0.50\pm0.07$ \\
$x_1$ &$-0.27\pm0.04$ & $-0.19\pm0.04$ \\
$a_3$ &$0.53\pm0.24$ & $0.32\pm0.15$ \\
$x_2$ &$0.85\pm0.81$ & $0.36\pm0.33$ \\
\hline
   \end{tabular}
\end{center}
\end{table} 

\begin{figure*}
\resizebox{\hsize}{!}{\includegraphics[angle=0]{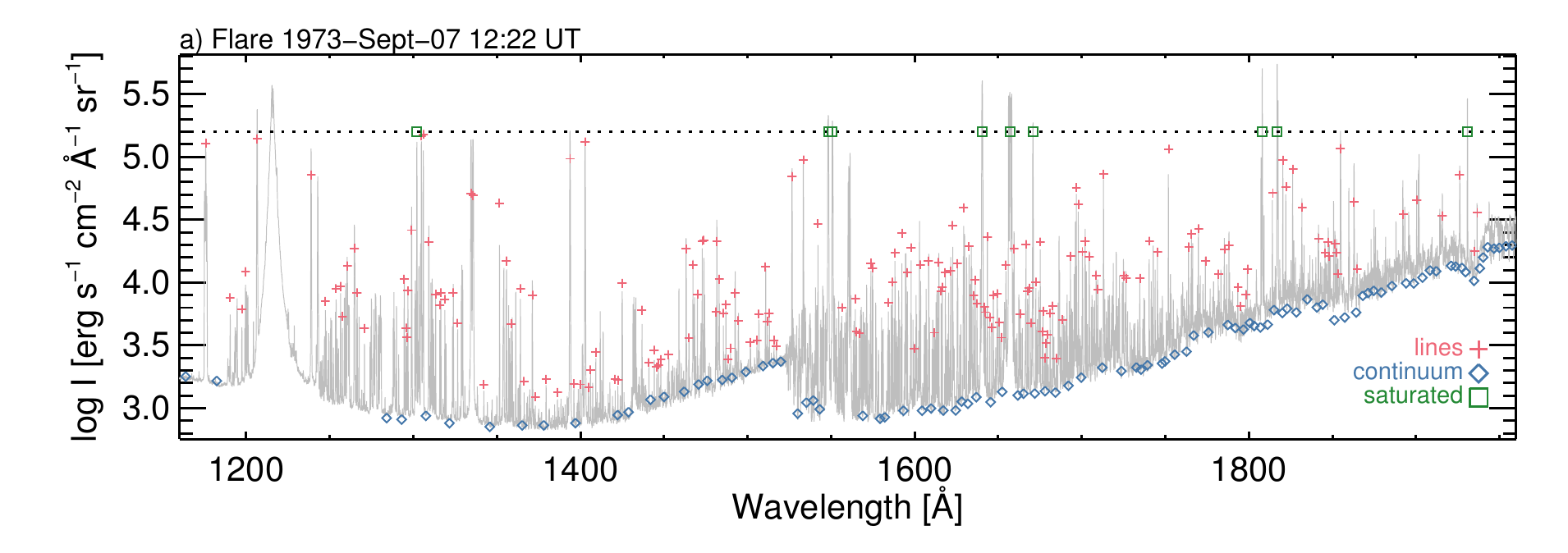}}
\resizebox{\hsize}{!}{\includegraphics[angle=0]{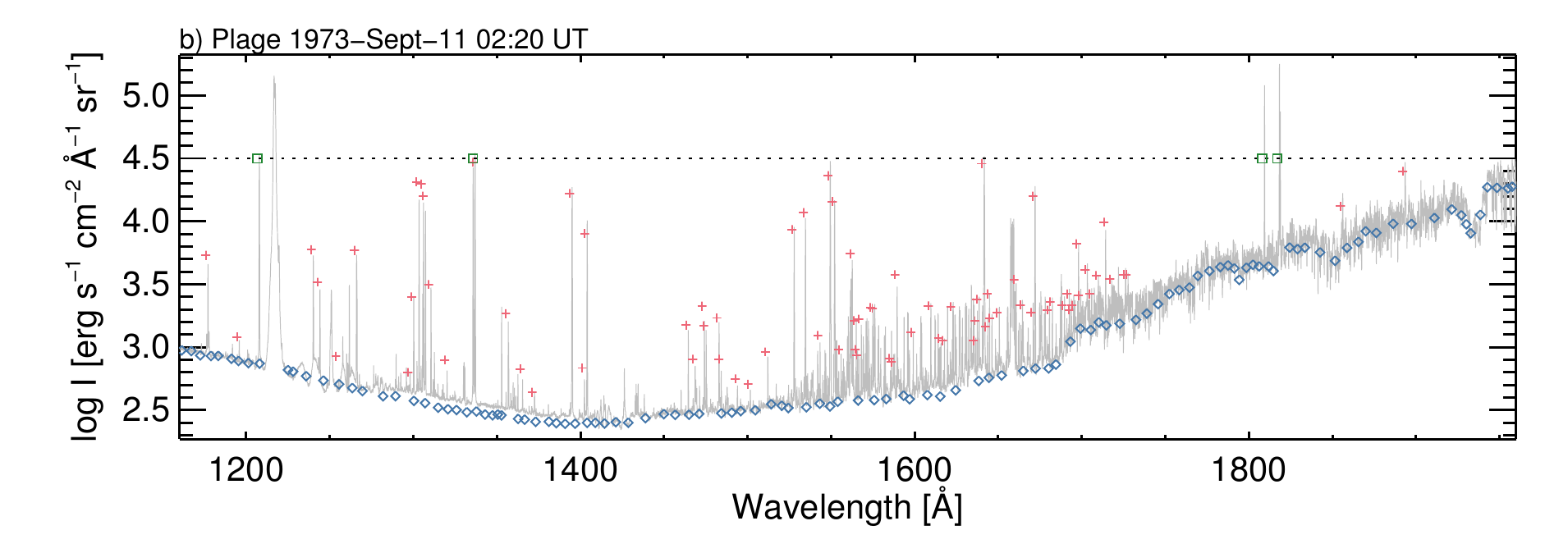}}
\caption{Calibrated NRL SO82B spectra for the (a) flare and (b) plage. The symbols (described in the legend) indicate digitized values obtained from \cite{DoyleCook:1992} and the dotted line shows the lower limit of the strongest lines, also from \cite{DoyleCook:1992}.}
\label{fig:calibflare}
\end{figure*}

\begin{figure}
\resizebox{\hsize}{!}{\includegraphics[angle=0]{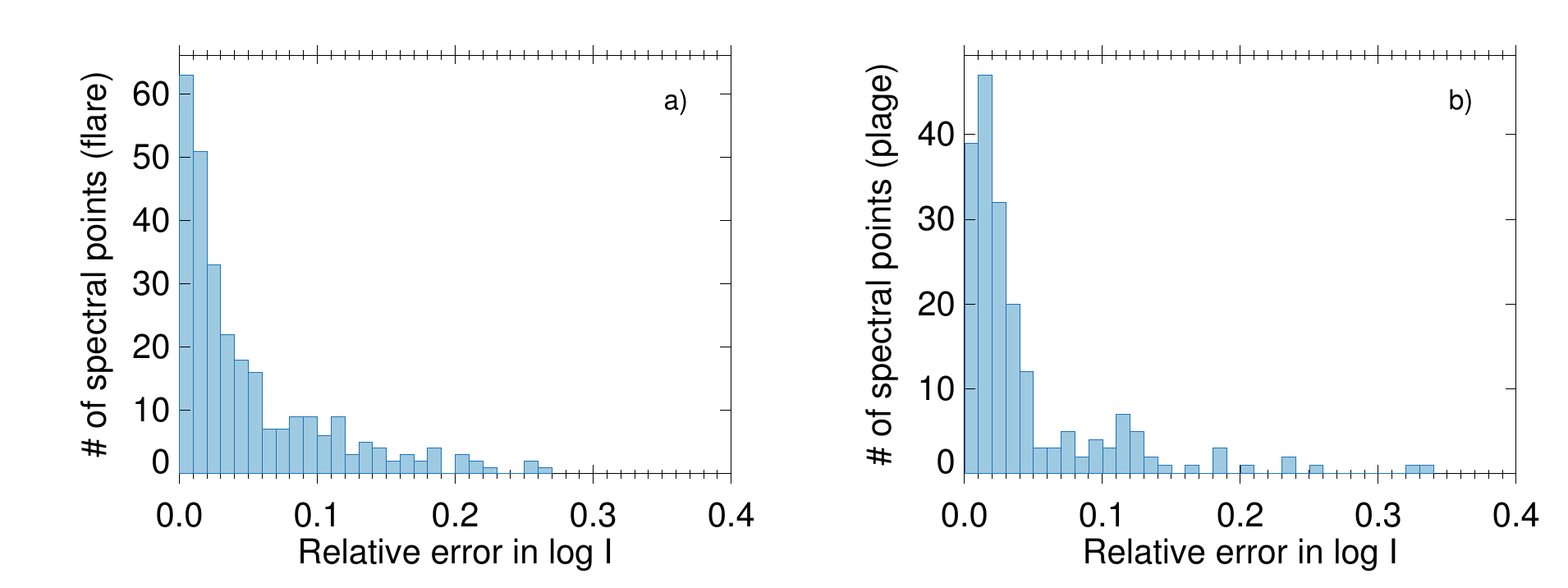}}
\caption{Distribution of the relative errors (in $\log I$) between digitized values from \cite{DoyleCook:1992} and our calibration for the (a) flare and (b) plage.}
\label{fig:err}
\end{figure}

Our final spectra, in specific intensity units, were obtained by using our characteristic curves (Figure~\ref{fig:hdfinal}), and show a good agreement with the spectra in \cite{DoyleCook:1992}, { as shown in Fig.~\ref{fig:calibflare} where the digitized values from \cite{DoyleCook:1992}, identified by the red, blue and green symbols, are compared with our calibrated spectra (in gray)}. The distribution of the relative errors (in $\log I$) between our calibrated values and the ones digitized from \cite{DoyleCook:1992} are shown in Figure~\ref{fig:err}, and are generally smaller than 15\%.

\bibliographystyle{aasjournal}
\bibliography{refs}

\end{document}